\documentclass[journal]{IEEEtran}

\usepackage{float}
\usepackage{epsf}
\usepackage{graphicx}
\DeclareGraphicsExtensions{.png}
\usepackage{caption}
\usepackage{subcaption}
\usepackage{xcolor}
\usepackage{amsmath} \interdisplaylinepenalty=2500
\usepackage{amsfonts}
\usepackage{amssymb}
\usepackage{array}
\usepackage{cuted}

\newcommand{\ul}[1]{\underline{#1}}

\newcommand{\eps}{\epsilon}

\newcommand{\sinc}{{\rm sinc}}

\newcommand{\bea}{\begin{eqnarray}}

\newcommand{\eea}{\end{eqnarray}}

\newcommand{\rmc}{{\rm c}}

\newcommand{\rmd}{{\rm d}}

\newcommand{\rmi}{{\rm i}}

\newcommand{\rmj}{{\rm j}}

\newcommand{\rmS}{{\rm S}}

\newcommand{\cB}{{\cal B}}

\newcommand{\cE}{{\cal E}}

\newcommand{\cF}{{\cal F}}

\newcommand{\cT}{{\cal T}}

\begin{document}



\title{Correlation and Spectral Density Functions in Mode-Stirred Reverberation -- I. Theory
}

\author{
{Luk R. Arnaut
}\\
}


\maketitle


\begin{abstract}
Auto- and cross-spectral density functions for dynamic {random} fields and power are derived. These are based on first- and second-order Pad\'{e} approximants of correlation functions expanded in terms of spectral moments. 
The second-order approximant permits a characterization of stir noise observable {at high stir frequencies in the autospectral density}. 
A relationship between stir imperfection and spectral kurtosis 
is established. For the latter, lower bounds are established. 
A novel alternative measure of correlation time for mean-square differentiable fields is introduced as the lag at the first point of inflection in the autocorrelation function. 
A hierarchy of Pad\'{e} deviation coefficients is constructed that quantify imperfections of correlations and spectra with increasing accuracy and range of lags. 
Analytical models of the spectral densities are derived and their asymptotic behaviour is analyzed.
The theoretical spectral density for the electric field (or $\rmS_{21}$) as an input quantity is compared with that for power (or $|\rmS_{21}|^2$) as the measurand. 
For the latter, its inverted-S shape conforms to experimentally observed stir-spectral power densities. The effect of additive noise on the stir autocorrelation and spectral density functions is quantified.
\end{abstract}



\section{Introduction\label{sec:intro}}
Autocorrelation functions (ACFs) play a central role in the characterization of second-order linear statistical dependence in random electromagnetic (EM) fields. In {the context of} mode-tuned reverberation chambers (MTRCs), for wide-sense stationary (WSS) swept fields with lags $\tau$, the ACF $\rho(\tau)$, its moments, and the correlation length $\tau_0$ express the efficiency of the field randomization process. 
This is often represented (but oversimplified) by an equivalent number of independent samples (degrees of freedom) in applications to extreme value statistics and uncertainty quantification of fields. 
This strategy focuses on withholding only uncorrelated states and discarding strongly correlated ones. It has been widely adopted to make the mode tuning process and EMC testing more efficient 
(cf., \cite{arnalocavg}--\cite{mons2015} and references therein).
For continuous stir processes, i.e., sweep traces in mode-stirred reverberation chambers (MSRCs), there is also interest in the neighborhood of extreme values, crossings and excursions above or below a set threshold level \cite{arnaexcessletter}, \cite{arnathresh}. 
Their estimation requires a precise functional characterization of $\rho(\tau)$, particularly for small lags ($\tau \ll\tau_0$) \cite{arnalocavg}, \cite{arnathresh}--\cite{caro2022}.
This perspective advocates oversampling in order to accurately characterize correlation at small lags, without discarding any serially correlated sampled data.

For spatially random fields in free space or near an EM boundary, analytic expressions for spatial ACFs have been derived \cite{hill1995}--\cite{arnaPRE}.
For stirred fields, the determination of ACFs has mainly relied on empirical models based on measured data \cite{arnalocavg}, \cite{arnaexcess} or reduced-order time series models \cite{lemo2008}.
Characterization of serial nonlinear and/or multipoint dependence (higher-order correlation) requires more general statistical methods \cite{arnacopula}.

From a different perspective, the spectral density function (SDF) of a stir process is related to its ACF via Fourier transformation, based on the Einstein--Wiener--Khinchin (EWK) theorem \cite{arnalocavg}, \cite{xu2018}. The SDF provides an alternative characterization of the stir process, in terms of periodicities (stir frequencies), offering additional insight about rates of field fluctuation across stir states. 
This is relevant in performing diagnostics, e.g., for a specific stirrer or drive mechanism, optimization of stirrer design and performance (unstirred energy), filtering of jitter, noise or harmonic EMI. To this end, the canonical stir spectrum for an ideal stirred chamber needs to be established, serving as a baseline for evaluating spectral distortions.

In \cite{arnalocavg}, stir spectral properties of an MTRC were investigated based on scalar measurements using an EM power sensor, at a single CW frequency. Analysis and modelling of such scalar data relied on an implicit assumption of circularity of the complex EM field.  When a  sampling vector network or spectrum analyzer (VNA/VSA) is available instead, the stir-spectral characterization in MSRCs can be refined and extended to greater accuracy in approaching a continuous stir trace \cite{arnathresh}, \cite{arnapulsejitt}, i.e., a time-continuous sweep of the local field $E(\tau)$ representing a sample path in stir space. This is a further motivation for the present work, revisiting \cite{arnalocavg}.

In this article, a nonparametric approach to the analytical modelling of ACFs, {cross-correlation functions} (CCFs) and SDFs of a stir process is taken.
The technique does not require an I/O model driven by a fictitious or extraneous source of random noise unlike, e.g., in autoregressive (AR) models \cite{lemo2008}. Instead, it is based on spectral moments that can be estimated from the stir sweep itself.
It is shown that the simple model proposed in \cite{arnalocavg} constitutes a zeroth-order approximation that can be extended consistently to higher orders by including higher-order spectral moments \cite{arnathresh}.
Secondly, whereas the local approximation of the ACF based on Taylor series expansion is accurate for very small lags, a Pad\'{e} approximation of equal order is shown to exhibit a larger range of approximation accuracy. 

Throughout the text, single- and double-primed quantities refer to real and imaginary parts of a complex quantity, respectively; typically originating from a WSS \cite{wong1985} complex electric field $E=E^\prime + \rmj E^{\prime\prime}$. For economy, $E^{\prime(\prime)}$ combines both cases $E^\prime$ and $E^{\prime\prime}$ in a single notation.
For the spectral moments $\lambda_m$ of SDFs, $\lambda^{\prime(\prime)}_m$ relate to $\rho^{\prime(\prime)}_E$ and $g^{\prime(\prime)}_E$, {\em and\/} already incorporate normalization by $\lambda_0\equiv \sigma^2_E$, i.e., $\lambda^{\prime(\prime)}_m \equiv \lambda_m / \lambda_0$.
An $\exp(\rmj\varpi\tau)$ dependence of the field is assumed and suppressed.

\section{Spectral Expansions of Random Fields}
\subsection{Spatial Random Fields}
In a static overmoded cavity, the angular plane-wave expansion of a random EM field $\ul{E}$ at a location $\ul{r}$ is given by
\begin{align}
\ul{E} ( \ul{r} ) = \frac{1}{|K|} \iiint_K \ul{\cE}(\ul{k}) \exp ( -\rmj \ul{k} \cdot \ul{r} ) \rmd\ul{k}
\label{eq:expansion_r}
\end{align}
where $\ul{r}$ and $\ul{k}=\omega \sqrt{\mu\eps}\, \ul{1}_k$ form a Fourier pair of conjugate spatial variables, and $|K|$ is the volume of the spherical $k$-shell of integration between radii $|\ul{k}_1|$ and $|\ul{k}_2|$. 
For a narrowband WSS field with spatial bandwidth $\Delta \ul{k} = |\ul{k}_1-\ul{k}_2| \ll |\ul{k}|=k$, the triple integration in (\ref{eq:expansion_r}) reduces to a double integration over $\ul{1}_k$ and solid angle $\Omega$ (surface of a spherical sector of unit radius), yielding the expansion
\begin{align}
\ul{E} ( \ul{r} ) = \frac{1}{|\Omega|} \iint_\Omega \ul{\cE}(\Omega) \exp ( -\rmj k \ul{1}_k \cdot \ul{r} ) \rmd\Omega.\label{eq:expansion_1r}
\end{align}
Spatially WSS fields $\ul{E}$ with respect to $||\ul{r}||$ and $\ul{1}_r$ (i.e., statistical spatial homogeneity and isotropy) exhibit delta-correlated covariance and correlation functions of $|\ul{k}_1-\ul{k}_2|$ \cite{wong1985}.

\subsection{Stirred Random Fields}
In a dynamic cavity, a random $\ul{E}$ is generated across an ensemble of statistically equivalent cavities, e.g., through stir states of one cavity that maintain equal interior volume, surface area and edge length throughout the stir process, to ensure a constant average mode density for statistical equivalence (equipartition). 
At an arbitrary interior location $\ul{r}_0$, the local field at a (continuous) stir state\footnote{For multistirring, an extension to a $d$-D stir vector $\ul{\tau}$ can be made \cite{arnathresh}.} $\tau$ can then be expanded as 
\begin{align}
\ul{E} (\tau) = \int^{+\infty}_{-\infty} \ul{\cE}(\varpi) \exp (\rmj\varpi \tau ) \rmd \varpi
\label{eq:expansion_tau}
\end{align}
where $\ul{\cE}(\varpi)$ is the stir process.
Only a scalar component $E_{x}$ will be considered, thereby dropping the subscript.
In continuous real-time stirring, stir traces form continuous sample paths in stir space and $\tau$ represents stir time $t$ in [$0,\cT$]. 

The angular frequency in the stir spectral domain, $\varpi$, is typically much lower than the CW excitation frequency, $\omega = 2\pi f$ \cite{arnaexcess}. 
Analogous to (\ref{eq:expansion_1r}), interest is here in narrowband stirring, i.e., relatively slow temporal variations of the field envelope (instantaneous amplitude of $E(\tau)$). 
Conversely, $\cE(\varpi) $ is the complex spectral density of ${E}(\tau)$ as its Fourier transform
\begin{align}
{\cE}(\varpi) = \int^{+\infty}_{-\infty} {E} (\tau) \exp (-\rmj\varpi \tau ) \rmd \tau
.
\label{eq:expansion_tau2}
\end{align}

For a bilinear product of fields, the spectral expansion is
\begin{align}
&{E}(\tau_1) {E}^{(*)}(\tau_2) = \nonumber\\
&\iint^{+\infty}_{-\infty} {\cE}(\varpi_1) {\cE}^{(*)}(\varpi_2) \exp [-\rmj(\varpi_1 \tau_1 \mp \varpi_2 \tau_2) ] \rmd \varpi_1 \rmd \varpi_2
\end{align} 
in which upper and lower signs refer to ${E}(\tau_1) {E}^{*}(\tau_2)$ and ${E}(\tau_1) {E}(\tau_2)$, respectively.
Ensemble averaging across all stir states results in the second-order (pseudo-)moment of $\ul{E}(\tau)$. For a WSS $\varpi$-differentiable stir process\footnote{The reference value of {$\langle |\cE^\varpi_0|^2 \rangle $} for $\cE(\varpi)$ is not necessarily the same as {$\langle |\cE^k_0|^2 \rangle$} for ${\cE}(\ul{k})$. Similarly, the functional expression for $\langle \cE(\varpi_1) \cE^{(*)}(\varpi_2) \rangle$ does not necessarily coincide with that for $\langle \cE(\ul{k}_1) \cE^{(*)}(\ul{k}_2) \rangle$.} and Kronecker $\delta(\cdot)$
\begin{align}
\langle \cE(\varpi_1) \cE^*(\varpi_2) \rangle &= \langle |\cE^{\varpi}_0|^2 \rangle \delta(\varpi_1-\varpi_2)\label{eq:cov_omega}\\
\langle \cE(\varpi_1) \cE(\varpi_2) \rangle &= 0
\label{eq:pcov_omega}
\\
\rho_E(\tau_1,\tau_2) &\stackrel{\Delta}{=} 
\frac{\langle E(\tau_1) E^*(\tau_2) \rangle}{\sqrt{\langle |E(\tau_1)|^2 \rangle \langle |E (\tau_2)|^2} \rangle} \nonumber\\
&= \int^{+\infty}_{-\infty} g_E(\varpi) \exp [\rmj\varpi (\tau_1 - \tau_2) ] \rmd \varpi
\label{eq:WK}
\end{align}
for arbitrary $\tau_1$,
with ${\langle |{E}(\tau_1)|^2} \rangle = {\langle|{E}(\tau_2)|^2} \rangle \stackrel{\Delta}{=} \sigma^2_E$ for assumed ${\langle {E}(\tau_1)} \rangle = {\langle{E}(\tau_2)} \rangle = 0$, where $\tau_1-\tau_2 \stackrel{\Delta}{=} \tau$ and
\begin{align} 
g_E(\varpi_1 - \varpi_2) = \frac{\langle \cE(\varpi_1) \cE^*(\varpi_2) \rangle}{\sigma^2_{E}}  \delta(\varpi_1-\varpi_2)
\label{eq:def_g_E}
\end{align}
is the normalized spectral density function (SDF) of $E(\tau)$.
Eq. (\ref{eq:WK}) shows that $\rho_E(\tau_1-\tau_2)$ and $g_E(\varpi_1 - \varpi_2)$ form a Fourier conjugate pair and as such is a statement of the Einstein--Wiener--Khinchin (EWK) theorem.

For noncircular $E(\tau)$, the pseudo-ACF $\langle E(\tau_1) E(\tau_2) \rangle / \sigma^2_E$  is the inverse Fourier transform of the pseudo-SDF
$\langle \cE(\varpi_1) \cE(\varpi_2) \rangle / \sigma^2_E$ in (\ref{eq:pcov_omega}) and extends the concept of the pseudo-correlation coefficient for $\tau=0$ \cite{arnaellipt} to nonzero lags in non-WSS fields. 
Such nonstationary fields may point at nonuniformity of stir speed, inhomogeneous sampling, nonequilibrium, etc.
For elliptic fields, the ACF and pseudo-ACF permit a complete second-order characterization of $E(\tau)$.

In summary, (\ref{eq:expansion_r}) and (\ref{eq:expansion_tau}) represent separate marginal field expansions of $\ul{E}(\ul{r},\tau)$,
as a function of $\ul{r}$ for arbitrary $\tau_0$ and as a function of $\tau$ for arbitrary $\ul{r}_0$, respectively.
The expectations, $\langle \cE(\ul{k}) \rangle$ and $\langle \cE(\ul{\varpi}) \rangle$, and the (pseudo-)covariances, $\langle \cE(\ul{k}_1) \cE^{(*)}(\ul{k}_2) \rangle_K$ and $\langle \cE(\varpi_1) \cE^{(*)}(\varpi_2) \rangle$ of the respective spectral processes are in general different.
 
\section{ACF and CCF of Stationary Complex $E$}
\subsection{General Case}
The ACF $\rho^\prime_E(\tau)$ and crosscorrelation function (CCF) $\rho^{\prime\prime}_E(\tau)$ of a WSS complex $E(t)$ can be expressed as
\begin{align}
&\rho_E(\tau) \equiv \rho^\prime_E (\tau) + \rmj \rho^{\prime\prime}_E(\tau) \stackrel{\Delta}{=} \frac{\langle E(0) E^*(\tau) \rangle}{\sigma^2_E}
\label{eq:rho_gen}\\
&= 
\frac{\sigma_{E^\prime,E^\prime}(0,\tau) + \sigma_{E^{\prime\prime},E^{\prime\prime}}(0,\tau)}{\sigma^2_{E^\prime} + \sigma^2_{E^{\prime\prime}}} + \rmj \frac{2\sigma_{E^{\prime\prime},E^\prime}(0,\tau)}{\sigma^2_{E^\prime} + \sigma^2_{E^{\prime\prime}}}
.
\label{eq:rho_gen_wss}
\end{align}
Eq. (\ref{eq:rho_gen_wss}) holds for general WSS fields ($\sigma^2_{E^{\prime(\prime)}}(0) = \sigma^2_{E^{\prime(\prime)}}(\tau) \stackrel{\Delta}{=} \sigma^2_{E^{\prime(\prime)}}$), 
noting that 
$\sigma_{E^{\prime\prime},E^\prime}(0,\tau) = \sigma_{E^{\prime},E^{\prime\prime}}(\tau,0) = - \sigma_{E^{\prime},E^{\prime\prime}}(0,\tau)$. For ideal circular $E$, we have that $\sigma^2_{E^\prime} = \sigma^2_{E^{\prime\prime}}$ and 
$\sigma_{E^\prime,E^{\prime\prime}}(0,\tau) = 0$, whence (\ref{eq:rho_gen})--(\ref{eq:rho_gen_wss}) simplify to 
\begin{align}
\rho_E(\tau)= \rho^{\prime}_E(\tau)= [\rho_{E^\prime}(\tau) + \rho_{E^{\prime\prime}}(\tau)]/2 = \rho_{E^{\prime(\prime)}}(\tau) 
.
\label{eq:rho_gen_circ}
\end{align}
For quasi-circular $E$, provided\footnote{Unlike for $\sigma_{E^\prime,E^{\prime\prime}}$ and $\sigma_{E^{\prime\prime},E^{\prime}}$, this condition between $\rho_{E^\prime,E^{\prime\prime}}$ and $\rho_{E^{\prime\prime},E^{\prime}}$ is not guaranteed to be satisfied because of the possibility of small different values of $\sigma_{E^\prime}$ and $\sigma_{E^{\prime\prime}}$ in the denominators of $\rho_{E^\prime,E^{\prime\prime}}$ and $\rho_{E^{\prime\prime},E^{\prime}}$.}
that
$\rho_{E^\prime,E^{\prime\prime}}(0,\tau) \simeq - \rho_{E^{\prime\prime},E^{\prime}}(0,\tau)$, eq. (\ref{eq:rho_gen_wss}) can be approximated as 
\begin{align}
\rho_E(\tau) 
&\simeq \frac{1}{2} \left \{ \rho_{E^\prime}(\tau) + \rho_{E^{\prime\prime}}(\tau) + \rmj \left [ \rho_{E^{\prime\prime},E^\prime}(\tau) - \rho_{E^\prime,E^{\prime\prime}} (\tau) \right ]\right \}
\nonumber\\
&\simeq \frac{1}{2} \left [ \rho_{E^\prime}(\tau) + \rho_{E^{\prime\prime}}(\tau) \right ] + \rmj \rho_{E^{\prime\prime},E^\prime} (0,\tau)\label{eq:rho_gen_approx}\\
&\simeq \rho_{E^{\prime(\prime)}}(\tau) + \rmj \rho_{E^{\prime\prime},E^\prime} (0,\tau)
.
\label{eq:rho_gen_simplified_approx}
\end{align}
Thus, (\ref{eq:rho_gen_wss}), (\ref{eq:rho_gen_circ}) and (\ref{eq:rho_gen_simplified_approx}) permit an interpretation of $\rho^\prime_E(\tau)$ and $\rho^{\prime\prime}_E(\tau)$ in terms of the in-phase/quadrature (I/Q) auto- and cross-covariance or -correlation functions for quasi-circular fields.
They further demonstrate that $\rho_E(\tau)$ is Hermitean conjugate, i.e., $\rho_E(\tau) = \rho^*_E(-\tau)$.

\subsection{Taylor Series Spectral Expansions}
\subsubsection{ACF}
For WSS ideal random fields, the I/Q CCF vanishes for any $\tau$, and the ACF is a real even function, i.e., $\rho_E(\tau) = \rho^\prime_E (-\tau)$. Its functional form can be obtained from spectral expansion, as follows.
Substituting the Fourier series $E^\prime(\tau)= \sum_i E_i \cos(\varpi_i \tau + \varphi_i)$ and $E^{\prime\prime}(\tau)= \sum_i E_i \sin(\varpi_i \tau + \varphi_i)$ into (\ref{eq:rho_gen}), {followed by} transitioning to a continuum of frequencies (i.e., $\Delta \varpi = 2\pi/N \rightarrow \rmd \varpi$ for $N \rightarrow +\infty$) yields
\begin{align}
{\sigma_{E^{\prime},E^{\prime}}(0,\tau)} = 
{\sigma_{E^{\prime\prime},E^{\prime\prime}}(0,\tau)} = 
{\sigma^2_E} \int^{+\infty}_0 g^\prime_E(\varpi) \cos(\varpi \tau) \rmd \varpi \label{eq:sigmaEpEp}
.
\end{align}
Applying a Taylor series expansion for $\cos(\varpi \tau)$, eq. (\ref{eq:sigmaEpEp}) results in a spectral expansion of $\sigma_{E^{\prime(\prime)},E^{\prime(\prime)}}(0,\tau)$ as
\begin{align}
&\sigma_{E^\prime, E^{\prime}}(0,\tau) = \sigma_{E^{\prime\prime}E^{\prime\prime}}(0,\tau) \nonumber\\
&= \sigma^2_{E^{\prime(\prime)}} (1 - \lambda^\prime_2 \tau^2 / 2! + \lambda^\prime_4 \tau^4 / 4! - \ldots),\hspace{2mm} |\tau| \leq \tau^\prime_c
\label{eq:acvf_expanded_Taylor}
\end{align}
and hence yields the spectral expansion of the ACF as 
\begin{align}
\rho^\prime_E(\tau) = 1 - ({\lambda^\prime_2}/{2!}) \tau^2 + ({\lambda^\prime_4}/{4!}) \tau^4 - \ldots, \hspace{2mm} |\tau| \leq \tau^\prime_c
\label{eq:rho_expanded_Taylor}
.
\end{align}
in which the radius of the region of convergence, $\tau^\prime_c$, depends on the moments $\lambda^\prime_{2i}$, where $i=1,2,\ldots$

Imposing the condition of mean-square differentiability of $E(\tau)$ \cite{wong1985} implies that $\dot{\rho}^\prime_E(\tau=0)=0$ \cite{vanm1983} and leads to the absence of a term proportional to  $|\tau|$ in (\ref{eq:rho_expanded_Taylor}) \cite{kac1959}.
This condition excludes certain idealized ACF models, e.g., $\rho_E(\tau) = \exp (- \tau/\tau_0)$ for the Ornstein--Uhlenbeck process relying on ideal white noise, but can be extended to discrete (mode tuning) processes by analytic continuation for $\tau \rightarrow 0$.   

\subsubsection{CCF}
In a similar way, the imaginary part $\rho^{\prime\prime}_E(\tau) = \rho_{E^{\prime\prime},E^\prime}(0,\tau)$ can be expressed in terms of the odd-order spectral moments of $E(\tau)$. 
Substituting the above Fourier series expansions of $E^\prime(\tau)$ and $E^{\prime\prime}(\tau)$ yields 
\begin{align}
{\sigma_{E^\prime,E^{\prime\prime}}(0,\tau)} = {\sigma^2_E} \int^{+\infty}_0 g^{\prime\prime}_E(\varpi) \sin(\varpi \tau) \rmd \varpi 
.
\label{eq: sigmaEpEpp}
\end{align}
For sufficiently small $\Delta \varpi$, the $E^{\prime}(\tau)$ and $E^{\prime\prime}(\tau)$ form a conjugate Hilbert pair with $E^{\prime\prime}(\tau) = \dot{E}^{\prime} (\tau) \equiv\rmd E^{\prime} (\tau) / \rmd \tau = \rmj \varpi E^{\prime} (\tau)$. A Taylor expansion of $\sin(\varpi \tau)$ in (\ref{eq: sigmaEpEpp}) results in
\begin{align}
&\sigma_{E^{\prime\prime},E^\prime}(0,\tau) = - \sigma_{E^\prime,E^{\prime\prime}}(0,\tau) \equiv \sigma_{\dot{E}^\prime,{E}^{\prime}}(0,\tau) \nonumber\\
&= \sigma^2_{E^{\prime(\prime)}} (\lambda^{\prime\prime}_1 \tau - \lambda^{\prime\prime}_3 \tau^3 / 3! +  \lambda^{\prime\prime}_5 \tau^5 / 5! - \ldots) , \hspace{2mm}|\tau| \leq \tau^{\prime\prime}_c
\end{align}
whence 
\begin{align}
\rho^{\prime\prime}_E(\tau) = \lambda^{\prime\prime}_1 \tau - ({\lambda^{\prime\prime}_3}/{3!}) \tau^3 + (\lambda^{\prime\prime}_5 / 5! ) \tau^5 -  \ldots , \hspace{2mm} |\tau| \leq \tau^{\prime\prime}_c
.
\label{eq:rhopp_expanded_Taylor}
\end{align}
If $E(\tau)$ is purely real then $\rho^{\prime\prime}_E(\tau)=0$, $g^{\prime\prime}_E(\varpi)=0$, $g_E(\varpi)=g^{\prime}_E(\varpi)$, and $\lambda^{\prime\prime}_{2i+1}=0$.

Since the complex CF and SDF form a Fourier transform pair (generalized EWK theorem; cf. sec. \ref{eq:genEWK}), these results are commensurate with general relationships between Fourier transforms, the derivative at $\tau=0$ and the first spectral moment \cite{brac1986}. 
Since $\dot{\rho}^\prime_E(0)/\rho^\prime_E(0) = \rmj \lambda^{\prime\prime}_1$ and $2\sigma^2_{E^{\prime(\prime)}} = \sigma^2_E$, it follows that $\dot{\rho}^\prime_E(0) = 0$ and $\dot{\rho}^{\prime\prime}_E(0) = \lambda^{\prime\prime}_1$. 

As shown in part II, odd-order spectral moments can be computed in the stir domain, without recourse to $g_E(\varpi)$, based on the CCF between $E(\tau)$ and its odd-order derivatives, e.g.,
\begin{align}
\sigma_{E^\prime,\dot{E}^{\prime\prime}} (t,t) = - \sigma_{\dot{E}^{\prime},{E}^{\prime\prime}} (t,t) = \lambda^{\prime\prime}_1 \equiv \sigma^2_E \int^{+\infty}_0 \varpi g^{\prime\prime}_E(\varpi) \rmd \varpi
.
\end{align} 
For numerical evaluation through finite-differenced sampled $E(\tau)$, higher-order spectral moments become increasingly sensitive to noise. In practice, this limits the order of expansion and spectral moments that can be extracted accurately. These sampling and noise effects on $\lambda^{\prime(\prime)}_m$ are analyzed in part II.

\subsection{Pad\'{e} Approximants for Series Expansion of ACF}
For experimentally determined CFs, the accuracy of the truncated Taylor series within its interval of convergence may be severely limited, restricted   
to a narrow range in the immediate vicinity of $\tau = 0$. As demonstrated in part III, 
an alternative Pad\'{e} approximation \cite{bake1996} offers better overall convergence over a wider range of $\tau$ {and larger interval of ordinates $\rho_E$}. This range increases with increasing order of the series.

The $(0,1)$-order (in $\tau^2$) Pad\'{e} approximant of $\rho^\prime_E(\tau)$ is
\begin{align} 
{\tilde\rho}^\prime_E(\tau^2) \equiv [0/1]_{\rho^\prime_E} (\tau^2) &\stackrel{\Delta}{=} \left ( 1+ a^\prime \tau^2 \right )^{-1}\nonumber\\
&= 1 - a^\prime \tau^2 + \left ( a^\prime \tau^2 \right )^2 - \ldots
\label{eq:rho_expanded_Pade_1storder}
\end{align}
showing that the linear (in $\tau^2$) model of this Pad\'{e} denominator, inspired by (\ref{eq:rho_expanded_Taylor}), permits a nonlinear model for $\tilde{\rho}^\prime_{E}(\tau^2)$ overall, with an infinite number of expansion terms.
Identifying equal powers of $\tau^2$ in (\ref{eq:rho_expanded_Taylor}) and (\ref{eq:rho_expanded_Pade_1storder}) to first order in $\tau^2$ shows that
\begin{align}
a^\prime = \lambda^\prime_2 /2.
\label{eq:aprime_Pade01}
\end{align}
For higher-order power terms, further identification yields $\lambda^\prime_4 = 6 (\lambda^\prime_2)^2$, $\lambda^\prime_6 = 90 (\lambda^\prime_2)^3$, etc., in this $(0,1)$-order approximation.
For ideal stirring (i.e., circular complex $E$), this Pad\'{e} approximant offers a precise model of the ACF for $|\tau| \leq \tau^\prime_c$ through just a single parameter $\lambda^\prime_2$, as clarified in sec. \ref{sec:ASDFfirstorderIdeal}. 

Unlike the first-order Taylor approximation $1 - a^\prime \tau^2$, eq. (\ref{eq:rho_expanded_Pade_1storder}) contains points of inflection at $\tau = \pm\tau^\prime_\rmi$ where the decay of the correlation switches from accelerating (i.e., concavity of $\ddot{\tilde\rho}^\prime_E(|\tau| < \tau^\prime_\rmi) < 0$) to decelerating (i.e., convexity of $\ddot {\tilde\rho}^\prime_E(|\tau| > \tau^\prime_\rmi) > 0$) as $|\tau|$ increases. 
The deceleration is essential in order that $\lim_{|\tau| \rightarrow +\infty} \rho^\prime_E(|\tau|) = 0$ and it enables $\tilde\rho^\prime_E(\tau)$ to track the empirical $\rho^\prime_E(\tau)$ for longer. 
Thus, $\tau^\prime_\rmi$ is a natural measure of correlation length $\tau_0$ and is given, in this first-order model, by 
\begin{align}
\tau^\prime_\rmi = \ddot{\tilde\rho}^{\prime^{-1}}_E(0) = (3a^\prime)^{-1/2} = (3\lambda^\prime_2/2)^{-1/2}
.
\end{align}

The $(0,2)$-order Pad\'{e} approximant in $\tau^2$ is the biquadratic
\begin{align}
[0/2]_{\rho^\prime_E} (\tau^2) &= \left ( 1 + a^\prime\tau^2 + b^\prime \tau^4 \right )^{-1}.
\label{eq:rhoE_expanded_Pade_2ndorder}
\end{align}
Termwise identification between the Taylor series for (\ref{eq:rhoE_expanded_Pade_2ndorder})
and ${\rho^\prime_E} (\tau^2)\simeq 1 - (\lambda^\prime_2/2!) \tau^2 + (\lambda^\prime_4/4!) \tau^4$ 
results in
\begin{align}
a^\prime = {\lambda^\prime_2} / {2},\hspace{3mm} b^\prime = - ( {\lambda^\prime_2} / {2} )^2 \kappa^\prime 
\end{align}
where the second-order deviation coefficient $\kappa^\prime$ is 
\begin{align}
{\kappa^\prime} \equiv - \frac{b^\prime}{(a^\prime)^2}
= \frac{(\lambda^\prime_4/4!) - (\lambda^\prime_2/2!)^2}{(\lambda^\prime_2/2!)^2} = \frac{\lambda^\prime_4}{6 (\lambda^\prime_2)^2} - 1
.
\label{eq:def_kappa}
\end{align} 
With reference to the corresponding double-sided SDF, this $\kappa^\prime$ measures the deviation of the spectral kurtosis from its value for ideal stirring (cf. sec. \ref{sec:ASDFfirstorderIdeal}).
For smooth continuous stir traces, $\kappa^\prime \geq -5/6$ because $\lambda^\prime_0\equiv 1$ and $(\lambda^\prime_2)^2 \leq \lambda^\prime_4$ on account of the Cauchy--Schwarz inequality \cite{cram1946}
\begin{align}
\left [ \int^{\infty}_0 \varpi^2 g^\prime_E(\varpi) \rmd\varpi \right ]^2 
\leq 
\int^{\infty}_0 g^\prime_E(\varpi) \rmd\varpi 
\int^{\infty}_0 \varpi^4 g^\prime_E(\varpi) \rmd\varpi 
.
\label{eq:CauchySchwarz}
\end{align}
The lower bound, $\kappa^\prime = -5/6$, is reached for either a purely deterministic field or a random phase field, i.e., $E(\tau)= a_0 \exp[\rmj(\varpi_0\tau+\phi)]$ with fixed $a_0$ and $\varpi_0$ but fixed or uniform $\phi$, for which $\cE(\varpi) = a_0 \exp(\rmj\phi) \delta(\varpi-\varpi_0)$ and $g^\prime_E(\varpi) \propto \cE(\varpi) \cE^*(\varpi) = |a_0|^2 \delta (\varpi-\varpi_0)$. 
Application of (\ref{eq:CauchySchwarz}) then yields $\lambda^\prime_4 = (\lambda^\prime_2)^2 = 3 |a_0|^4$. 
Alternatively, representing a deterministic field as a Gaussian field with zero variance, an application of Isserlis's theorem yields the same result. 
By extension, a value of $\kappa^\prime$ between $-5/6$ and $0$ may indicate a partially randomized field, i.e., a superposition of an ideal circular random (stirred) field (random $a$ and/or $\varpi$) and a deterministic (unstirred) field. 
As shown in part II, a negative $\kappa^\prime$ can also arise as an artefact of finite-difference sampling.

In summary, a Pad\'{e} approximation provides a `pull-up' in the ACF model beyond its first point of inflection, offering a better approximation to the experimental ACF across a wider interval of $\tau$, compared to a Taylor series of equal order.
In the $(0,2)$-order approximant, ${\kappa^\prime}$ serves as a spectral measure of stir imperfection as stir-spectral kurtosis.
Examples and results for different orders of Taylor and Pad\'{e} approximations for ACFs of measured data will be given in part III.

\section{Field Spectral Density Functions\label{sec:FSDF}}
\subsection{Terminology}
In this article, a slight departure is made from commonlu used terminology that associates second-order SDFs with ``power'' (or energy) being the SDF {\em output\/} quantity. 
For reasons to be clarified in sec. \ref{sec:PSDF}, it is advantageous in MSRCs to be able to determine the SDF also from $|E|^2$ -- rather than just $E$ -- as the {\em input\/} quantity (measurand).
Therefore, ``field SDFs'' (FSDFs) here refer to SDFs that take $E$ as input quantity (output of a VNA or VSA), which is conventionally denoted as the (power) SDF, reserving the term ``power SDFs'' for SDFs that assume $|E|^2$ (or energy, or power) as the SDF input quantity (output of a power meter). 
For FSDFs, further distinction is made between the auto-FSDF (ASDF) and cross-FSDF (CSDF). For power SDFs, the cross-SDF is zero, owing to real $|E|^2$, whence the PSDF is always an auto-PSDF.  

\subsection{Auto- and Cross-Spectral Density Functions\label{eq:genEWK}}
In (\ref{eq: sigmaEpEpp}), the imaginary part $g^{\prime\prime}_E(\varpi) \stackrel{\Delta}{=} {\cal F}[\rho^{\prime\prime}_E(\tau)](\varpi)$ represents the I/Q-based CSDF (quadrature cross-SDF) associated with the quadrature field CCF.
The generalized EWK theorem follows as the Fourier transform relationship
\begin{align}
g_E(\varpi) \equiv g^\prime_E(\varpi) + \rmj g^{\prime\prime}_E(\varpi) \stackrel{\Delta}{=} \int^{+\infty}_{-\infty} \rho_E(\tau) \exp(\rmj \varpi \tau) \rmd \tau
.
\label{eq:complexEWK}
\end{align}
This holds on account of the even symmetry of the ACF of  $E^{\prime(\prime)}$ and odd symmetry of the CCF between $E^\prime$ and $E^{\prime\prime}$,
resulting in a separation between the purely real ASDF and purely imaginary CSDF (quadrature cross-spectrum) because $E^\prime(\tau)$ and $E^{\prime\prime}(\tau)$ differ in phase by $\pi/2$. 
Therefore, the ASDF and CSDF as real and imaginary parts of $g_E(\varpi)$ are related to the ACF and CCF via
\begin{align}
g^\prime_E(\varpi) &= \int^{+\infty}_{-\infty} \left [ \rho^\prime_E(\tau) \cos(\varpi \tau) - \rho^{\prime\prime}_E(\tau) \sin(\varpi \tau) \right ] \rmd \tau \nonumber\\
&\rightarrow \int^{+\infty}_{-\infty} \rho^\prime_E(\tau) \cos(\varpi \tau) \rmd \tau \label{eq:ASDF}
\\
g^{\prime\prime}_E(\varpi) &= \int^{+\infty}_{-\infty} \left [ \rho^{\prime\prime}_E(\tau) \cos(\varpi \tau) + \rho^\prime_E(\tau) \sin(\varpi \tau) \right ] \rmd \tau 
.
\label{eq:CSDF}
\end{align}
The asymptotic expression in (\ref{eq:ASDF}) contains only contributions by the ACF. This limit is reached for ideal stirring, where $\rho^\prime_E(\tau)=\delta(\tau)$ (Kronecker).

As an alternative to the EWK theorem, which relies on obtaining the entire ACF and CCF first, the SDF, ASDF and CSDF can also be obtained directly from the Fourier transform of the data-based periodogram (for finite data sets), viz.,
\begin{align}
G_E(\varpi) 
&= \langle \left (\cE^\prime(\varpi) \right )^2 \rangle
+ \langle \left (\cE^{\prime\prime}(\varpi) \right )^2 \rangle 
+ \rmj 2 \langle \cE^\prime(\varpi) \cE^{\prime\prime}(\varpi) \rangle
\label{eq:periodgram}
\end{align}
where $\cE(\varpi) \stackrel{\Delta}{=} \cF[E(\tau)](\varpi)$ and $\langle \cE^\prime(\varpi) \cE^{\prime\prime}(\varpi) \rangle \rightarrow 0$.
Since the periodogram is generally more efficiently computed than the covariance matrix for sampled data, an alternative computation of $\rho_E(\tau)$ is offered by (\ref{eq:periodgram}) followed by inverse transformation of (\ref{eq:complexEWK}), i.e.,
\begin{align}
E(\tau) \stackrel{{\rm FT}}{\longrightarrow} 
\cE(\varpi) \stackrel{(\ref{eq:periodgram})}{\longrightarrow} 
g_{E}(\varpi )\stackrel{{\rm IFT}}{\longrightarrow} 
\rho_E(\tau)
.
\label{eq:periodogram_cont}
\end{align}

\subsection{ASDF Based on Pad\'{e} Approximated ACF}
\subsubsection{First Order\label{sec:ASDFfirstorderIdeal}}
The ASDF model based on the $(0,1)$-order Pad\'{e} approximant (\ref{eq:rho_expanded_Pade_1storder}) is obtained by its Fourier {cosine} transformation. Its poles are
$z^\prime_0 = - z^\prime_1 = \rmj \sqrt{2/\lambda^\prime_2}$,
whence $g^\prime_E(\varpi)$ follows as
\begin{align}
\tilde{g}^\prime_E(\varpi) = \frac{1}{\pi} \cF_c \left ( \frac{{2}/{\lambda^\prime_2}}{\tau^2 + \frac{2}{\lambda^\prime_2}} \right ) (\varpi) 
= \sqrt{\frac{2}{\lambda^\prime_2}}
\exp\left ( -\sqrt{\frac{2}{\lambda^\prime_2}}\,\varpi \right )
\label{eq:psdf_exp}
\end{align}
valid for $\varpi > 0$. The corresponding  double-sided\footnote{Recall that spectral moments of even order can be expressed equivalently based on single- as well as double-sided ASDFs \cite{vanm1983}.} 
ASDF, $\tilde{s}^\prime_E(\varpi)= \tilde{g}^\prime_E(|\varpi|)/2$ for $-\infty < \varpi < +\infty$, is a Laplacean with mean $\lambda^\prime_1=0$ and kurtosis $\lambda^\prime_4/(\lambda^\prime_2)^2 = 6$, for which $\kappa^\prime=0$. Thus, $\kappa^\prime$ 
is the excess spectral kurtosis with reference to an ideal Laplacean ASDF, scaled by a factor $1/6$.

Eq. (\ref{eq:psdf_exp}) represents the ASDF for ideal stirred fields. 
A complex {circular} Gaussian probability distribution of $E(\tau)$ {transforms} to a {complex circular} Gaussian\footnote{{For a discrete time series of $N_s$ samples $E(\tau_n)$ with sufficiently rapidly decaying $p$-point correlations of any order $p\geq 2$, this result also holds asymptotically for its $N_s$-point discrete Fourier transform, i.e., asymptotic Gaussianity of $\cE(\varpi_k)$ for each $k=0,\ldots,N_s-1$ when $N_s\rightarrow +\infty$ \cite{bril1981}}.} $\cE(\varpi)$ with orthogonal frequency components.
Therefore, {for $\varpi\not=0$,} the non-normalized SDF ${G}_E(\varpi) \sigma^2_E g_E(\varpi)$ has a scaled $\chi^2_2$ distribution when the degrees of freedom tend to $+\infty$.

Departing from this limiting $\chi^2_2$ ASDF, when $\cE(\varpi)$ and $g^\prime_E(\varpi)$ are restricted to a finite stir bandwidth $\cB$, this affects the ACF and its Pad\'{e} approximants. 
In particular, a rectangular SDF $g^\prime_E(\varpi)=1/\cB$ corresponds to
\begin{align}
\rho^\prime_E(\tau)=\sinc(\cB\tau)
\label{eq:sinc_rhoE}
\end{align} 
approaching a unit impulse ACF in the limit $\cB\rightarrow +\infty$. 
For small $|\tau|$, (\ref{eq:sinc_rhoE}) remains quadratic to leading order, viz., $1-(\cB\tau)^2/6$, similarly to (\ref{eq:rho_expanded_Pade_1storder}), but now with an increased correlation length. 
The functional form (\ref{eq:sinc_rhoE}) with $kr$ replacing $\cB \tau$ is recognized as the familiar ideal spatial ACF \cite{hill1995}.

\subsubsection{Second Order}
\begin{figure}[htb] \begin{center}
\begin{tabular}{c}
\hspace{-0.6cm}
\includegraphics[scale=0.65]{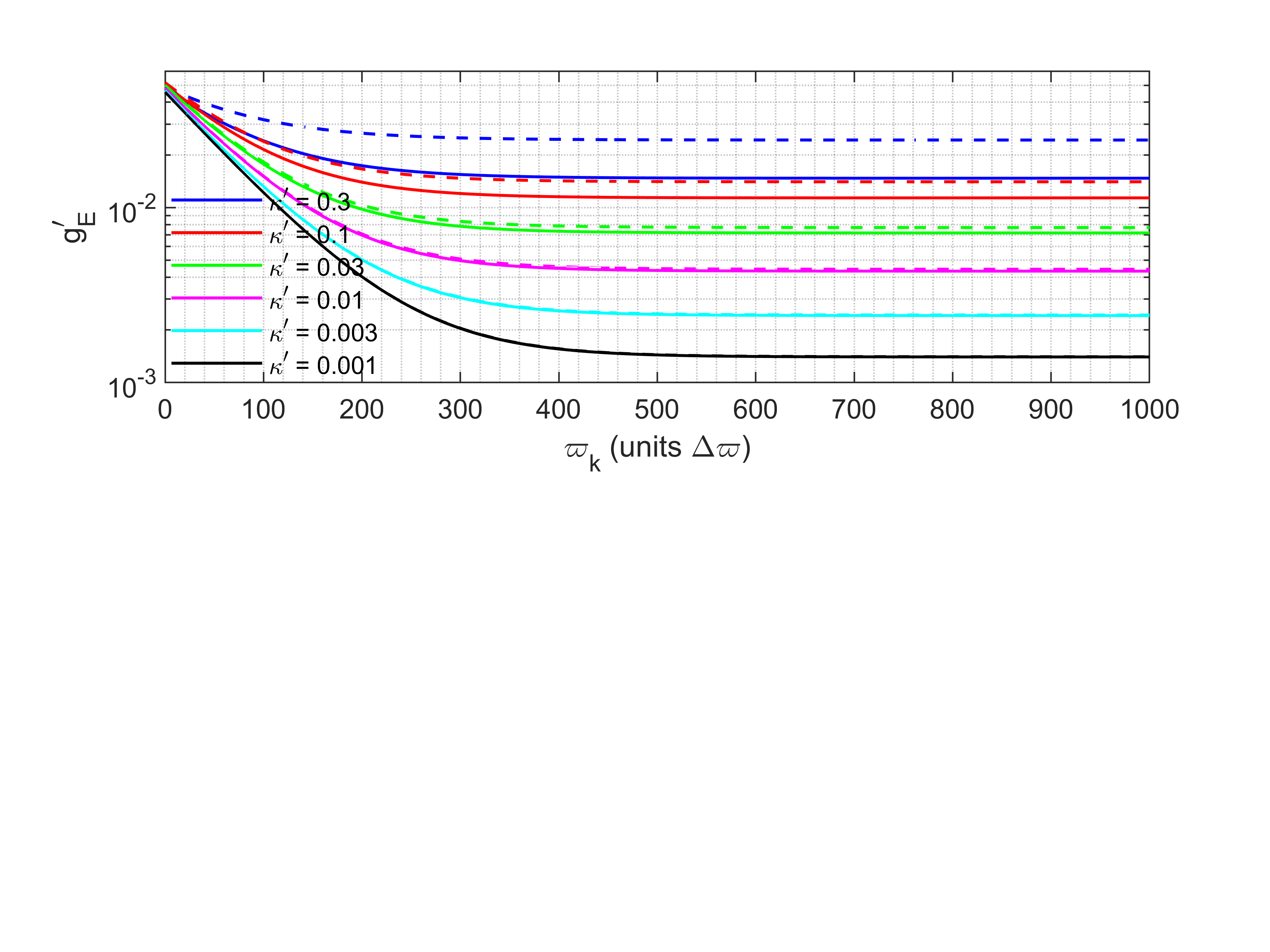}\\ 
\\
\vspace{-4.5cm}\\
\end{tabular}
\end{center}
{
\caption{\label{fig:ASDF_param_kappa}
\small
Envelopes of theoretical ASDFs $\tilde{g}^\prime_E(\varpi)$, based on 
general-$\kappa^\prime$ (solid) and approximate small-$\kappa^\prime$ (dashed) $[0/2]$-order Pad\'{e} approximant, at selected $\kappa^\prime$, for $(\lambda^\prime_2)^{1/2}=100$ rad/s and $\Delta \varpi=0.01$ rad/s.}
}
\end{figure}

For $\kappa^\prime > 0$, the squared poles of  (\ref{eq:rhoE_expanded_Pade_2ndorder}) are $z^{\prime 2}_{0,1} = (-a^\prime + s^\prime)/(2b^\prime)$ and $z^{\prime 2}_{2,3} = (-a^\prime - s^\prime)/(2b^\prime)$ where $s^\prime \stackrel{\Delta}{=} \sqrt{a^{\prime^2}-4b^\prime} = (\lambda^\prime_2/2) \sqrt{1+4{\kappa^\prime}} \geq a^\prime > 0$. One pole $z^\prime_0$ lies on the positive imaginary axis; two others are real, viz., 
\begin{align}
z^{\prime}_{0} = - z^{\prime}_{1} = \rmj \sqrt{\frac{s^{\prime}-a^{\prime}}{2|b^{\prime}|}},\hspace{3mm}
z^{\prime}_{2} = - z^{\prime}_{3} = \sqrt{\frac{s^{\prime}+a^{\prime}}{2|b^{\prime}|}}
.
\end{align}
Integration of 
$\psi^\prime_E(z) \stackrel{\Delta}{=} \tilde{\rho}^{\prime}_E(z) \exp(\rmj \varpi z)/\pi$ 
along a contour that is closed in the upper half of the $z$-plane ($z\stackrel{\Delta}{=}\tau+\rmj \xi$, $\xi \geq 0$) by a semi-circle enclosing $z^\prime_0$ yields, upon application of the residue and Cauchy limit theorems in $z^{\prime}_{2}$ and $z^{\prime}_{3}$
\begin{align}
\tilde{g}^{\prime}_E(\varpi) 
&= \lim_{\epsilon_2,\epsilon_3\rightarrow 0} \left ( \int^{z^{\prime}_{3}-\epsilon_3}_{-\infty} + \int^{z^{\prime}_{2}-\epsilon_2}_{z^{\prime}_{3}+\epsilon_3} + \int^{+\infty}_{z^{\prime}_{2}+\epsilon_2} \right ) \psi^\prime_E(\tau) \rmd \tau \nonumber\\
&=\frac{\exp \left ( - |z^\prime_{0}| \varpi \right )}{s^\prime|z^\prime_{0}|} + \frac{\sin \left ( z^\prime_{2}\varpi \right )}{s^\prime z^\prime_{2}},\hspace{2mm}\varpi > 0
.
\label{eq:SDFcaseII}
\end{align}
Fig. \ref{fig:ASDF_param_kappa} shows the envelope of $\tilde{g}^\prime_E(\varpi)$ in (\ref{eq:SDFcaseII}) for selected $\kappa^\prime$.
When $0<\kappa^\prime \ll 1/4$, i.e., for $0<4|b^\prime|/{a^\prime}^2 \ll 1$, 
with 
$(s^\prime - a^\prime)/(2|b^\prime|) = (\sqrt{1+4{\kappa^\prime}} - 1)/(\kappa^\prime \lambda^\prime_2 ) 
\simeq 2(1-\kappa^\prime)/\lambda^\prime_2$
and
$(s^\prime + a^\prime)/(2|b^\prime|) \simeq 2/(\kappa^\prime \lambda^\prime_2)$,
eq. (\ref{eq:SDFcaseII}) becomes explicitly
\begin{align}
\tilde{g}^\prime_E(\varpi)
&\simeq 
\sqrt{\frac{2}{\lambda^\prime_2}} 
\left ( 1 - \frac{3\,\kappa^\prime}{2} \right ) \exp \left ( - \sqrt{(1-{\kappa^\prime})\frac{2}{\lambda^\prime_2}} \, \varpi \right ) 
\nonumber\\ 
&\hspace{2mm}
+ 
\sqrt{\frac{2 \kappa^\prime}{\lambda^\prime_2}} 
\sin \left ( \sqrt{\frac{2}{{\kappa^\prime} \lambda^\prime_2 }} \,\varpi \right )
.
\label{eq:SDFcaseII_smallkappa}
\end{align}
The second term in (\ref{eq:SDFcaseII}) and (\ref{eq:SDFcaseII_smallkappa}) is governed by the real poles $\tau_{2} = - \tau_3 \simeq \sqrt{2/( {\kappa^\prime} \lambda^\prime_2 )}$ that emerge when extending the order of the Pad\'{e} approximation for $\rho^\prime_E(\tau)$ from $[0/1]$ to $[0/2]$. 
For $0 < \kappa^\prime \ll 1/4$, these poles are situated far from the origin, at locations where $\rho^\prime_E(\tau_{2})$ is dominated by correlation noise and thus exhibiting rapid fluctuations with $\tau$. 
In other words, the second term in (\ref{eq:SDFcaseII}) and (\ref{eq:SDFcaseII_smallkappa}) represents HF stir noise with a relatively small spectral amplitude $\sqrt{2\kappa^\prime / \lambda^\prime_2}$ compared to the exponentially decaying first term that dominates near stir DC. 
Recall that (\ref{eq:SDFcaseII}) is a functional approximation to the actual $g^\prime_E(\varpi)$ because it is derived from the $[0/2]$-order approximant
(\ref{eq:rhoE_expanded_Pade_2ndorder}). 
For $\varpi \ll \varpi^\prime_c$ or $\varpi \gg \varpi^\prime_c$ (cf. (\ref{eq:cornerfreq_ASDF})), the envelope of (\ref{eq:SDFcaseII_smallkappa}) is approximately
\begin{align}
&\tilde{g}^\prime_E(\varpi)
\simeq \nonumber\\
&\sqrt{\frac{2}{\lambda^\prime_2}}
\left [ 
\left ( 1 - \frac{3\,\kappa^\prime}{2} \right ) \exp \left ( - \sqrt{(1-{\kappa^\prime})\frac{2}{\lambda^\prime_2}} \, \varpi \right ) 
+ \sqrt{\kappa^\prime} \right ]
\label{eq:SDFcaseII_smallkappa_env}
\end{align}
which is also shown in Fig. \ref{fig:ASDF_param_kappa} for comparison.

When further higher-order terms as powers of $\tau^2$ are included in the expansion of (\ref{eq:rhoE_expanded_Pade_2ndorder}), more pairs of poles arise that contribute additional terms to $\tilde{g}^\prime_E(\varpi)$. This gives rise to higher-order deviation coefficients for the departures of the ACF from ideal stirring at large lags, to $m$th order in $\tau^2$. These coefficient are 
\begin{align}
\kappa^\prime_{m} \stackrel{\Delta}{=}
\frac{{\lambda^\prime_{2m}}/{(2m)!} - \left (  {\lambda^\prime_2}/{2!} \right )^m}{\left (  {\lambda^\prime_2}/{2!} \right )^m}
= \frac{2^{m}\lambda^\prime_{2m}}{(2m)!\hspace{1mm} (\lambda^\prime_2)^{m}}-1
\label{eq:ACFdevcff_kappap}
\end{align} 
for $m\geq 0$, including $\kappa^\prime_0=\kappa^\prime_1=0$ for vanishing zeroth- and first-order deviations owing to $\lambda^\prime_0 \equiv 1$, as well as $\kappa^\prime_2 \equiv \kappa^\prime$. Since these $\kappa^\prime_m$ are associated with increasing powers of the delay $\tau$ as $m$ increases, they constitute a hierarchy of ACF and SDF deviations of increasing order as a function of $\tau$. Such deviation coefficients are useful for relatively long-range correlation, e.g., at relatively low CW frequencies, inefficient stirring, high EM absorption, etc. 

Higher-order terms in $\tilde{g}^\prime_E(\varpi)$ perturb the stir DC value and the stir LF rate of decay of $\tilde{g}^\prime_E(\varpi)$ when $\kappa^\prime_m \not = 0$. 
They also produce a progressively finer structure in the SDF, superimposed onto the dominant exponential first term in (\ref{eq:SDFcaseII}).
This finer detail becomes significant only at exceedingly small and large $|\varpi|$, when the exponential term that decays at a rate $\sqrt{2(1-{\kappa^\prime}) /\lambda^\prime_2}$, has become sufficiently small. 

\subsubsection{Asymptotic Behavior}
Several asymptotic results can be deduced from the approximants, as follows.
\paragraph{Levels}
\begin{itemize}
\item Near stir DC ($\varpi \rightarrow 0+$), the first term in (\ref{eq:SDFcaseII}) approaches
\begin{align}
\tilde{g}^\prime_E(\varpi \rightarrow 0+) 
&\simeq 
\frac{1 - |z^\prime_0| \varpi }{s^\prime |z^\prime_0|} 
\nonumber\\
&\simeq
\sqrt{\frac{2}{ \lambda^\prime_2}} \left ( 1 - \frac{3}{2} {\kappa^\prime} \right )
-
\frac{2\pi}{\left ( 1 + 2 {\kappa^\prime} \right ) \lambda^\prime_2}
\varpi
.
\end{align}
Formally extended to $\varpi=0$ itself, its value is lower for realistic stirring ($\kappa^\prime > 0$) than for ideal stirring ($\kappa^\prime = 0$).
\item For continuous stir traces and efficient stirring ($0 < \kappa^\prime \ll 1/4$), 
the envelope associated with the dominant second term in (\ref{eq:SDFcaseII}) in the $[0/2]$-order approximation for $\varpi \rightarrow +\infty$ approaches the asymptotic value
\begin{align}
|\tilde{g}^\prime_E(\varpi \rightarrow +\infty)| = \left( z^\prime_{2} s^\prime \right )^{-1}
\simeq \sqrt{{2 \kappa^\prime}/{\lambda^\prime_2}} 
.
\label{eq:ASDF_maxfreq_cont}
\end{align}
Consequently, the relative level drop of the SDF from its level near stir DC ($\varpi=0+$) to that at $\varpi\rightarrow+\infty$ is then
\begin{align}
\frac{\tilde{g}^\prime_E(+\infty)}{\tilde{g}^\prime_E(0+)} &= \frac{|z^\prime_0|}{ z^\prime_{2}} = \sqrt{\frac{s^\prime-a^\prime}{s^\prime+a^\prime}} 
\simeq 
\sqrt{\kappa^\prime}
\label{eq:leveldrop_cont}
\end{align} 
indicating that the explicit dependence of both $\tilde{g}^\prime_E(+\infty)$ and $\tilde{g}^\prime_E(0+)$ on $\sqrt{\lambda^\prime_2}$ cancels in their ratio, in this second-order approximation. 
Thus, a nonzero HF level of the ASDF represents stir imperfection (stir noise $\kappa^\prime \not = 0$). By contrast, in the first-order approximation, i.e., ideal stirring, $\tilde{g}^\prime_E(+\infty)/\tilde{g}^\prime_E(0+)$ approaches zero on account of $g^\prime_E(\varpi) \propto \exp(-|z^\prime_0|\varpi)$.
\item A ``corner'' stir frequency $\varpi^\prime_c$ for the ASDF can be defined, where the LF exponential decay of $g^\prime_E(\varpi)$ transitions to HF stir noise. Equating the first term of (\ref{eq:SDFcaseII}) and the envelope of the second term in magnitude yields
\begin{align}
\varpi^\prime_c &= 
\sqrt{\frac{2 |b^\prime|}{s^\prime-a^\prime}} 
{\rm ln} \left (
\sqrt{\frac{s^\prime + a^\prime}{s^\prime-a^\prime}} 
\right )
\simeq 
- \sqrt{\frac{\lambda^\prime_2}{8(1-{\kappa^\prime})}}
{\rm ln} ( \kappa^\prime )
.
\label{eq:cornerfreq_ASDF}
\end{align}
This $\varpi^\prime_c$ defines a duration $\cT^\prime_c = 2\pi/\varpi^\prime_c\propto 1/\sqrt{\lambda^\prime_2}$ that can be interpreted as a stir autocorrelation time measure.
\end{itemize}

\paragraph{Rates}
\begin{itemize}
\item
In the $[0/1]$-order model (\ref{eq:psdf_exp}), the slope at $\varpi=0+$ is $-2/\lambda^\prime_2$. 
For the dominant first term in (\ref{eq:SDFcaseII}), the slope is  
\begin{align}
&\frac{\rmd \tilde{g}^{\prime}_E(\varpi\ll \varpi^\prime_c)}{\rmd \varpi} = 
- \frac{1}{s^\prime} \exp \left ( - \sqrt{\frac{s^\prime-a^\prime}{2|b^\prime|}} \, \varpi \right ) \label{eq:SDFcaseII_gen_slope}\\
&\simeq - \frac{2}{\lambda^\prime_2 }
\left (1-2{\kappa^\prime} \right ) \exp \left [ - \left (1-{\kappa^\prime} \right ) \sqrt{\frac{2}{\lambda^\prime_2 }} \varpi \right ]
.
\label{eq:SDFcaseII_gen_slope_smallkappa}
\end{align}
This indicates a slower decay with $\varpi$ for stir imperfections ($\kappa^\prime > 0$) compared to ideal stirring ($\kappa^\prime=0$).
\item For $\varpi \ll \varpi^\prime_c$ and $0 < \kappa^\prime \ll 1/4$, 
\begin{align}
\frac{\rmd \tilde{g}^\prime_E(\varpi \ll \varpi^\prime_c)}{\rmd \varpi} 
&= - ({s^\prime})^{-1} 
\simeq - {\frac{2}{\lambda^\prime_2}} \left ( 1 - 2 {\kappa^\prime} \right ) 
\end{align}
indicating that the rate of exponential decay diminishes as $\kappa^\prime$ increases from zero. 
For ideal stirring ($\kappa^\prime=0$), it follows again that ${\rmd \tilde{g}^\prime_E(0+)}/{\rmd \varpi} = -2/\lambda^\prime_2$.
\item For $\varpi\gg \varpi^\prime_c$, the slope of the envelope for the second term in (\ref{eq:SDFcaseII}) approaches
\begin{align}
\frac{\rmd \tilde{g}^\prime_E(\varpi \gg \varpi^\prime_c)}{\rmd \varpi} 
&= - \left ( z^\prime_{2} s^\prime \right )^{-1}
\simeq - \sqrt{{2 \kappa^\prime}/{\lambda^\prime_2}}
.
\end{align}
Since $\sqrt{\lambda^\prime_2}$ is proportional to the CW excitation frequency $f$ in the limit of continuous or alias-free sampled stir traces ($\Delta\tau\rightarrow 0$) (cf. parts II and III), it follows that, to first order in $\kappa^\prime$, the quasi-DC and HF parts of $\tilde{g}^\prime_E(\varpi)$ decay at rates proportional to $1/f^2$ and $1/f$, respectively.
\end{itemize}

The ASDF (\ref{eq:SDFcaseII}) for $\kappa^\prime > 0$ can be extended to the range $\kappa^\prime < 0$, as follows. Negative $\kappa^\prime$ arise, in particular, for sampled stir data when the sampling rate is relatively low or after data decimation (cf. part II). In such cases, stir HF content is captured with a reduced accuracy. Specifically, for $-1/4 \leq \kappa^\prime < 0$, a derivation similar to the foregoing leads to 
\begin{align}
&\tilde{g}^{\prime}_E(\varpi) 
= 
\frac{\sqrt{2 b^\prime}}{s^\prime} 
\left [
\frac{\exp \left ( -\sqrt{\frac{a^{\prime} - s^{\prime}}{2 b^{\prime}}} \varpi \right )}{\sqrt{a^\prime - s^{\prime}}}
-
\frac{\exp \left ( -\sqrt{\frac{a^{\prime} + s^{\prime}}{2 b^{\prime}}} \varpi \right )}{\sqrt{a^\prime + s^{\prime}}}
\right ]
\end{align}
which can be expressed for $-1/4 \ll \kappa^\prime < 0$ as
\begin{align}
\tilde{g}^{\prime}_E(\varpi) 
&\simeq
\frac{\sqrt{2/\lambda^\prime_2}}{1-2|\kappa^\prime|} \left [
\exp \left ( - \sqrt{\frac{2}{\lambda^\prime_2}} \varpi \right ) \right.
\nonumber\\
&\hspace{2mm} \left.
- 
\sqrt{\frac{|\kappa^\prime|}{1-|\kappa^\prime|}}
\exp \left ( - \sqrt{\frac{2}{\lambda^\prime_2} \frac{1-|\kappa^\prime|}{|\kappa^\prime|}} \varpi \right )
\right ]
.
\end{align}
For $-5/6 \leq \kappa^\prime \leq -1/4$, the ASDF is
\begin{align}
&\tilde{g}^{\prime}_E(\varpi) 
= 
\frac{\rmj}{(z^{\prime^2}_0-z^{\prime^2}_1) b^\prime} 
\left [
\frac{\exp \left ( \rmj {z^\prime_0} \varpi \right )}{{z^{\prime}_0} }
-
\frac{\exp \left ( \rmj {z^\prime_1} \varpi \right )}{{z^{\prime}_1} }
\right ]
\label{eq:gep_negkappa_lessthan_oneoverfour}\\
&= 
\sqrt{
\frac{|\kappa^\prime|}{|\kappa^\prime|-1/4}} \exp (-y^\prime \varpi) \left [ x^\prime \cos (x^\prime \varpi) + y^\prime \sin (x^\prime \varpi) \right ]
\end{align}
where ${z^\prime_{0,1}} \equiv \pm x^\prime + \rmj y^\prime$ are the poles in the upper half-plane
\begin{align}
z^{\prime}_{0,1} &= \sqrt{\frac{\sqrt{{a^\prime}^2 + {r^\prime}^2}}{2 b^\prime}} \exp \left \{ \frac{\rmj}{2} \left [ \tan^{-1} \left (\mp \frac{r^\prime}{a^\prime} \right ) + \pi \right ] \right \} \\
x^\prime &\stackrel{\Delta}{=} \sqrt{\frac{2}{\lambda^\prime_2 \sqrt{|\kappa^\prime|}}} \sin \left ( \frac{1}{2} \tan^{-1} \left (\sqrt{4|\kappa^\prime|-1}\right ) \right )\\
y^\prime &\stackrel{\Delta}{=} \sqrt{\frac{2}{\lambda^\prime_2 \sqrt{|\kappa^\prime|}}} \cos \left ( \frac{1}{2} \tan^{-1} \left (\sqrt{4|\kappa^\prime|-1} \right ) \right )
\end{align}
in which upper and lower signs refer to ${z^{\prime}_0}$ and ${z^{\prime}_1}$, respectively, $r^\prime \stackrel{\Delta}{=} \sqrt{4 b^\prime - {a^\prime}^2} = (\lambda^\prime_2 / 2)\sqrt{4 |\kappa^\prime|-1}$ and $(z^{\prime^2}_0-z^{\prime^2}_1) b^\prime = \rmj r^\prime$. 

\subsection{CSDF Based on Pad\'{e} Approximated CCF}
For the CSDF $g^{\prime\prime}_E(\tau)$, corresponding results are similarly obtained by Pad\'{e} approximation of the truncated expansion (\ref{eq:rhopp_expanded_Taylor}) for $\rho^{\prime\prime}_E(\tau)$, leading to the following results.
Normalization of a CSDF is a delicate issue. Therefore, the various CSDFs in this section are up to a normalizing multiplication constant.

\subsubsection{Zeroth Order}
A basic model for $g^{\prime\prime}_E(\varpi)$ can be obtained from a linear approximation\footnote{Strictly, (\ref{eq:gppE_Pade10}) is not a Pad\'{e} approximant, for its definition requires the denominator to be of minimum order one in $\tau$.} of the CCF for very small $\tau$, i.e.,
\begin{align}
[1/0]_{\rho^{\prime\prime}_E}(\tau) = \lambda^{\prime\prime}_1 \tau,\hspace{2mm}|\tau| \ll \tau^{\prime\prime}_c
\label{eq:gppE_Pade10}
\end{align} 
based on (\ref{eq:rhopp_expanded_Taylor}). Its Fourier transform
\begin{align}
\tilde{g}^{\prime\prime}_E(\varpi) =  \frac{\lambda^{\prime\prime}_1}{\varpi} 
\left [ \frac{\sin(\varpi\tau^{\prime\prime}_c)}{\varpi} - \tau^{\prime\prime}_c \cos (\varpi \tau^{\prime\prime}_c)
\right ]
\end{align}
may be used when $\tau^{\prime\prime}_c$ is known. 

\subsubsection{First Order}
The Pad\'{e} approximant of order $[1/2]$ in $\tau$ (i.e., with first-order denominator in $\tau^2$) for  (\ref{eq:rhopp_expanded_Taylor}) is 
\begin{align}
[1/2]_{\rho^{\prime\prime}_E}(\tau) = \frac{c^{\prime\prime} \tau}{1+a^{\prime\prime} \tau^2}
\label{eq:gppE_Pade12}
\end{align} 
where $a^{\prime\prime} = {\lambda^{\prime\prime}_3}/({6\lambda^{\prime\prime}_1})$ and $c^{\prime\prime}=\lambda^{\prime\prime}_1$ by identification. Fourier {sine} transformation of (\ref{eq:gppE_Pade12}) results in the first-order model
\begin{align}
\tilde{g}^{\prime\prime}_E(\varpi) 
&=
{\frac{6 (\lambda^{\prime\prime}_1)^2}{\lambda^{\prime\prime}_3}}
\exp\left ( -\sqrt{\frac{6 \lambda^{\prime\prime}_1}{\lambda^{\prime\prime}_3}}\,\varpi \right )
.
\label{eq:csdf_exp}
\end{align}
A formal analogy betttween (\ref{eq:csdf_exp}) and (\ref{eq:psdf_exp}) can be seen by noting that $\tilde{g}^\prime_E(\varpi)$ is governed by the stir time constant $\beta^{\prime} \stackrel{\Delta}{=} \sqrt{2! \lambda^\prime_0/\lambda^\prime_2}$, whereas $\tilde{g}^{\prime\prime}_E(\varpi)$ depends on $\beta^{\prime\prime} \stackrel{\Delta}{=} \sqrt{3!\lambda^{\prime\prime}_1/\lambda^{\prime\prime}_3}$.

\subsubsection{Second Order}
Starting from the second-order (in $\tau^2$) Taylor approximation of the CCF (\ref{eq:rhopp_expanded_Taylor}), i.e., 
\begin{align}
\rho^{\prime\prime}_E(\tau) \simeq \frac{\lambda^{\prime\prime}_1 \tau}{1!} \left ( 
1 
- \frac{\lambda^{\prime\prime}_3}{3!\lambda^{\prime\prime}_1} \tau^2 
+ \frac{\lambda^{\prime\prime}_5}{5!\lambda^{\prime\prime}_1} \tau^4
\right ), \hspace{2mm} |\tau| \leq \tau^{\prime\prime}_c
\label{eq:rhopp_expanded_Taylor_2ndorder_rewrite}
\end{align}
its Pad\'{e} approximant of order $[1/4]$ in $\tau$ is
\begin{align}
\tilde{\rho}^{\prime\prime}_E(\tau) \stackrel{\Delta}{=}
[1/4]_{\rho^{\prime\prime}_E} (\tau) &= \frac{c^{\prime\prime} \tau}{1 + a^{\prime\prime}\tau^2 + b^{\prime\prime} \tau^4}
.
\label{eq:rhopp_expanded_Pade_2ndorder_repeat}
\end{align}
Identification of (\ref{eq:rhopp_expanded_Taylor_2ndorder_rewrite}) with the Taylor expansion of (\ref{eq:rhopp_expanded_Pade_2ndorder_repeat}) up to order $\tau^5$ results in
\begin{align}
a^{\prime\prime} = \frac{\lambda^{\prime\prime}_3}{6\lambda^{\prime\prime}_1},\hspace{2mm}
b^{\prime\prime} = - \left ( \frac{\lambda^{\prime\prime}_3}{6\lambda^{\prime\prime}_1} \right )^2 \kappa^{\prime\prime}, \hspace{2mm}
c^{\prime\prime}= \lambda^{\prime\prime}_1
\end{align}
where, by the same token as (\ref{eq:def_kappa})--(\ref{eq:CauchySchwarz})
\begin{align}
\kappa^{\prime\prime} \equiv \kappa^{\prime\prime}_2 \stackrel{\Delta}{=} {\frac{3\lambda^{\prime\prime}_1\lambda^{\prime\prime}_5}{10 (\lambda^{\prime\prime}_3)^2} - 1}
 \end{align}
leading to $\kappa^{\prime\prime} \geq -7/10$ on account of $(\lambda^{\prime\prime}_3)^2 \leq \lambda^{\prime\prime}_1 \lambda^{\prime\prime}_5$.

Expressions for $\tilde{g}^{\prime\prime}_E(\varpi)$ follow along similar lines as for $\tilde{g}^{\prime}_E(\varpi)$, now by contour integration of
$\psi^{\prime\prime}_E(z) \stackrel{\Delta}{=}  \tilde{\rho}^{\prime\prime}_E(z) \exp(\rmj \varpi z) / \pi$.
For $\kappa^{\prime\prime} > 0$,
the poles of (\ref{eq:rhopp_expanded_Pade_2ndorder_repeat}) are
\begin{align}
z^{\prime\prime}_{0} = - z^{\prime\prime}_{1} = \rmj \sqrt{\frac{s^{\prime\prime}-a^{\prime\prime}}{2|b^{\prime\prime}|}}
,\hspace{2mm}
z^{\prime\prime}_{2} = - z^{\prime\prime}_{3} = \sqrt{\frac{s^{\prime\prime}+a^{\prime\prime}}{2|b^{\prime\prime}|}}
\end{align}
where $s^{\prime\prime} \stackrel{\Delta}{=} \sqrt{(a^{\prime\prime})^2-4b^{\prime\prime}} = \sqrt{1+4{\kappa^{\prime\prime}}} \lambda^{\prime\prime}_3 /(6\lambda^{\prime\prime}_1)$ 
and 
$(s^{\prime\prime}\pm a^{\prime\prime})/(2|b^{\prime\prime}|) = 3 (\sqrt{1+4{\kappa^{\prime\prime}}}\pm 1) \lambda^{\prime\prime}_1 / ( \kappa^{\prime\prime} \lambda^{\prime\prime}_3)$.
This results in 
\begin{align}
\tilde{g}^{\prime\prime}_E(\varpi) \textcolor{black}{=}  
\frac{\lambda^{\prime\prime}_1}{s^{\prime\prime}} \left [ \exp \left ( - |z^{\prime\prime}_{0}| \varpi \right ) - \cos \left ( z^{\prime\prime}_{2}\varpi \right ) \right ],\hspace{2mm}
\kappa^{\prime\prime}>0
.
\label{eq:SDFppcaseII}
\end{align}
Explicitly, for $0<\kappa^{\prime\prime} \ll 1/4$, it follows that
\begin{align}
\tilde{g}^{\prime\prime}_E(\varpi)
&\simeq \frac{6 (\lambda^{\prime\prime}_1)^2}{\left ( 1+2 {\kappa^{\prime\prime}} \right ) \lambda^{\prime\prime}_3}
\left [ 
\exp \left ( - \sqrt{\left ( 1-{\kappa^{\prime\prime}} \right )\frac{6\lambda^{\prime\prime}_1}{\lambda^{\prime\prime}_3} } \, \varpi \right ) 
\right .
\nonumber\\ 
&\hspace{27mm}\left . 
- 
\cos \left ( \sqrt{\frac{6\lambda^{\prime\prime}_1}{ {\kappa^{\prime\prime}} \lambda^{\prime\prime}_3}} \,\varpi \right ) 
\right ]
\label{eq:SDFcaseIIpp_smallkappa}
\end{align}
whose asymptotic envelope is
\begin{align}
\tilde{g}^{\prime\prime}_E(\varpi)
&\simeq \frac{6 (\lambda^{\prime\prime}_1)^2}{\left ( 1+2 {\kappa^{\prime\prime}} \right ) \lambda^{\prime\prime}_3}
\left [ 
\exp \left ( - \sqrt{\left ( 1-{\kappa^{\prime\prime}} \right )\frac{6\lambda^{\prime\prime}_1}{\lambda^{\prime\prime}_3} } \, \varpi \right ) + 1
\right ]
.
\label{eq:SDFcaseIIpp_smallkappa_env}
\end{align} 

\begin{figure}[htb] \begin{center}
\begin{tabular}{c}
\hspace{-0.6cm}
\includegraphics[scale=0.65]{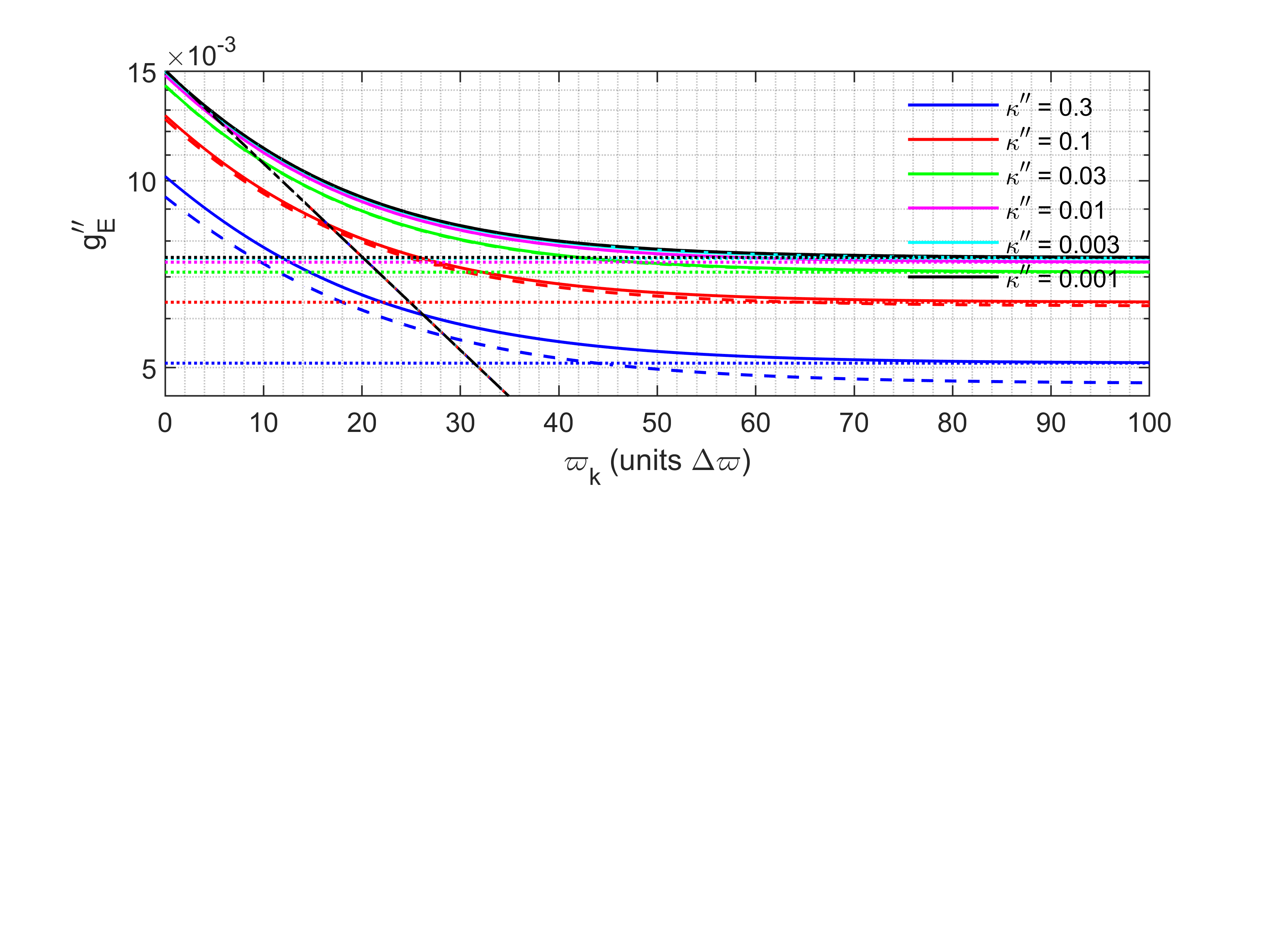}\\ 
\\
\vspace{-4.5cm}\\
\end{tabular}
\end{center}
\caption{\label{fig:CSDF_param_kappa}
\small
Envelopes of theoretical CSDFs: 
$[1/2]$-order (dot-dashed black), $[1/4]$-order  general-$\kappa^{\prime\prime}$ (solid), $[1/4]$-order small-$\kappa^{\prime\prime}$ approximative (dashed), and $[1/4]$-order HF asymptotic (dotted) Pad\'{e} approximant based $\tilde{g}^{\prime\prime}_E(\varpi)$ as a function of $\varpi_k=k\Delta \varpi$, at selected $\kappa^{\prime\prime}$ for $\lambda^{\prime\prime}_1=0.5$ rad/s, $(\lambda^{\prime\prime}_3)^{1/3}=8.55$ rad/s, $\Delta \varpi = 0.01$ rad/s.
}
\end{figure}

Fig. \ref{fig:CSDF_param_kappa} compares 
(\ref{eq:SDFcaseIIpp_smallkappa_env}) with
(\ref{eq:csdf_exp}) at selected values of $\kappa^{\prime\prime}> 0 $, $\lambda^{\prime\prime}_1$ and $\lambda^{\prime\prime}_3$. First- and second-order approximations are seen to merge when $\kappa^{\prime\prime} \rightarrow 0$, while the level drop $\tilde{g}^{\prime\prime}_E(\varpi \rightarrow +\infty) / \tilde{g}^{\prime\prime}_E(\varpi \rightarrow 0+)$ is relatively small compared to that for $\tilde{g}^\prime_E(\varpi)$.

For $-1/4 \leq \kappa^{\prime\prime} < 0$, all poles $z^{\prime\prime}_{i}$ are imaginary whence  
\begin{align}
\tilde{g}^{\prime\prime}_E(\varpi) 
&= 
\frac{c^{\prime\prime}}{a^{\prime\prime}} 
\left [
\exp \left ( -\sqrt{\frac{a^{\prime\prime} - s^{\prime\prime}}{2 b^{\prime\prime}}} \varpi \right )
\right. \nonumber\\ 
&\left. \hspace{12mm}
-
\exp \left ( -\sqrt{\frac{a^{\prime\prime} + s^{\prime\prime}}{2 b^{\prime\prime}}} \varpi \right )
\right ]\\ 
&\simeq
\frac{6 (\lambda^{\prime\prime}_1)^2}{\lambda^{\prime\prime}_3} \left [
\exp \left ( 
- \sqrt{\frac{6\lambda^{\prime\prime}_1}{\lambda^{\prime\prime}_3}} \varpi \right ) 
\right. \nonumber\\ 
&\left. \hspace{19mm}
- 
\exp \left ( 
- \sqrt{\frac{1-|\kappa^{\prime\prime}|}{|\kappa^{\prime\prime}|} \frac{6\lambda^{\prime\prime}_1}{\lambda^{\prime\prime}_3} } \varpi \right )
\right ]
\end{align}
where the latter approximation holds for $-1/4 \ll \kappa^{\prime\prime} < 0$.

In analogy with $\kappa^\prime_m$ in (\ref{eq:ACFdevcff_kappap}), deviation coefficients can be similarly defined for the CCF and CSDF. 
These are associated with higher-order departures of the $[1/4]$-order $\tilde{\rho}^{\prime\prime}_E(\tau)$ from the $[1/2]$-order approximant and are found as 
\begin{align}
\kappa^{\prime\prime}_{m} \stackrel{\Delta}{=}
\frac{\frac{\lambda^{\prime\prime}_{2m+1}}{(2m+1)!\lambda^{\prime\prime}_1} - \left ( \frac{\lambda^{\prime\prime}_3}{3!\lambda^{\prime\prime}_1}\right )^m}{\left ( \frac{\lambda^{\prime\prime}_3}{3!\lambda^{\prime\prime}_1}\right )^m}
= \frac{6^{m}\lambda^{\prime\prime}_{2m+1} (\lambda^{\prime\prime}_1)^{m-1}}{(2m+1)!\hspace{1mm} (\lambda^{\prime\prime}_3)^{m}}-1
\label{eq:CCFdevcff_kappapp}
\end{align} 
for $m\geq 0$, including $\kappa^{\prime\prime}_0=\kappa^{\prime\prime}_1=0$.
Unlike for the ACF, however, the $[1/2]$-order approximant does not represent ideal stirring, for which $\rho^{\prime\prime}_E(\tau) = 0$. Therefore, (\ref{eq:CCFdevcff_kappapp}) only measures departures with respect to this $[1/2]$ approximant.  

Corresponding results for asymptotic levels and slopes of the CSDF at low and high stir frequencies as for the ASDF can be derived {\it mutatis mutandis}, upon replacing single-primed quantities with corresponding double-primed counterparts. 

\subsubsection{CSDF vs. ASDF}
Comparing the CSDF (\ref{eq:csdf_exp}) and ASDF (\ref{eq:psdf_exp}) at low stir frequencies, the ratio of the cross- to autospectral density levels is
\begin{align}
\frac{g^{\prime\prime}_E(\varpi) }{ g^{\prime}_E(\varpi) } = \frac{6\sqrt{2\lambda^\prime_2}(\lambda^{\prime\prime}_1)^2 }{\lambda^{\prime\prime}_3}
\end{align} 
which is independent of the excitation frequency $f$, to leading order, on account of $(\lambda^\prime_m)^{1/m} \propto f$ for continuous stir traces.
Similarly, the ratio of their respective rates of LF decay, i.e., $\sqrt{3\lambda^{\prime\prime}_1 \lambda^\prime_2 / \lambda^{\prime\prime}_3}$, is again independent of $f$.

\section{Power Spectral Density Functions\label{sec:PSDF}}
\subsection{First-Order Approximation}
When stir sweep data derive from direct measurements of the scalar power $\rmc U$ using a power sensor, rather than the complex field $E$ using a VSA or complex $S_{21}$ using a VNA, the corresponding ACF $\rho_U$ differs from $\rho_E$. 
In analogy with the EWK theorem applied to $\rho_E$, the Fourier transform of $\rho_U$ can be taken and is denoted now as a power SDF (PSDF), $g_U(\varpi)$. 
In \cite{bell1965}, the transformation of spectral moments for various nonlinear (including square-law) envelope detectors with input $g_E(\varpi)$ was studied.
In \cite{arnaexcess}, the empirical best-fit model for the ACF of $U\propto |E|^2$, based on measured complex $E$, was found -- among several alternative candidate mean-square differentiable field models -- as
\begin{align}
\rho_U(\tau) \simeq  \left [ 1 + \left ( \frac{\tau}{2\tau_c} \right )^2 \right ]^{-2}
\label{eq:rhoU}
\end{align}
confirming experimental results in \cite{arnalocavg} obtained using a power sensor.
The model (\ref{eq:rhoU}) conforms to the theoretical relationship 
$\rho_U(\tau) \simeq \rho^2_E(\tau)$ for ideal random fields \cite{midd}, but with different correlation lengths and bandwidths.
It has the form
\begin{align}
\tilde{\rho}_U(\tau) \equiv [0/1]_{\rho_U} (\tau^2) = 
\left [ 1 + (\tau/\beta^{\prime}_U)^2 \right ]^{-2}
.
\end{align}
The first-order Pad\'{e} approximation for the PSDF follows by integration of 
\begin{align}
\psi_U(z) = \pi^{-1} \exp(\rmj \varpi z)/ [1 + (z/\beta^{\prime}_U)^2]^2
\end{align}
along a closed contour with a centered semicircular arc in the upper half-plane ($z=\tau + \rmj \xi$; $\xi \geq 0$),
enclosing the double pole at $z_0=\rmj \beta^{\prime}_U$ whose residue is
\begin{align}
R_{z_0} [\psi_U(z)] = - \rmj\beta^{\prime}_U
(1+\beta^{\prime}_U\varpi) \exp(-\beta^{\prime}_U\varpi)/(4\pi)
.
\end{align}
Application of Jordan's lemma along the arc whose radius is extended to infinity and using \cite[eq. (17.34.8)]{grad} now yields a polynomial-exponential PSDF model for $U$, viz.,
\begin{align} 
\tilde{g}_U(\varpi) &= {\pi}^{-1}\cF_c[\tilde{\rho}_U(\tau)](\varpi) 
= \int^{+\infty}_0 \frac{\cos(\varpi \tau)}{\pi [1+(\tau/\beta^{\prime}_U)^2]^2} \rmd \tau\nonumber\\
&= \beta^{\prime}_U \left ( 1+ \beta^{\prime}_U \varpi \right ) \exp \left ( -\beta^{\prime}_U\varpi \right ) / 2
\label{eq:psdf_nonexp_U}
\end{align}
in contrast to the pure exponential ASDF (\ref{eq:psdf_exp}) for $E$.
The approach of (\ref{eq:psdf_nonexp_U}) to stir DC is markedly different from that for the ASDF, because $\dot{\tilde{g}}_U(\varpi\rightarrow 0+)=0$ and $\ddot{\tilde{g}}_U(\varpi\rightarrow 0+)<0$, whereas $\dot{\tilde{g}}^\prime_E(\varpi\rightarrow 0+)<0$ and $\ddot{\tilde{g}}^\prime_E(\varpi\rightarrow 0+)>0$.

In contrast to the zeroth-order PSDF model \cite[eq. (7)]{arnalocavg}
\begin{align}
\tilde{g}_U(\varpi) = \beta^{\prime}_U\exp(-\beta^{\prime}_U\varpi)
\label{eq:psdf_nonexp_U,zerothorder}
\end{align}
the improving first-order model (\ref{eq:psdf_nonexp_U}) features spectral flattening at low stir frequencies, owing to the additional polynomial factor. 
This produces an increased stir-HF spectral density and a higher cut-off stir frequency. 
Physically, this flattening arises because $(|\cE(\varpi)|^2)^2$ for $g_U(\varpi)$ produces more extreme values than $|\cE(\varpi)|^2$ does for $g_E(\varpi)$, leading to an increased spectral density at HFs. 
This effect parallels the fact that the $\chi^2_4$ probability distribution of $|E|^4$ has a larger mean than the $\chi^2_2$ probability distribution of $|E|^2$.

\subsection{Second-Order Approximation\label{sec:2ndorder_gU}}
A second-order model for $g^\prime_U(\varpi)$, based on the $[0/2]$ Pad\'{e} approximant of $\rho_U$ in $\tau^2$, i.e., $\tilde{\rho}^\prime_U(\tau^2) = (1 + a^\prime \tau^2 + b^\prime \tau^4)^{-2}$, follows similarly as before, now by integration of
\begin{align}
\psi^\prime_U(z) = \pi^{-1} \exp(\rmj \varpi z) / (1 + a^\prime z^2 + b^\prime z^4)^2
\label{eq:rhoU_expanded_Pade_2ndorder}
\end{align}
along the same contour as used with $\psi^\prime_E(z)$. This results in
\begin{align}
&\tilde{g}^{\prime}_U(\varpi) \simeq \nonumber\\
&\frac{1}{2 {s^\prime}^2 |z^\prime_{0}|^3 } \left [ \left ( 1 + \frac{4|b^\prime||z^\prime_{0}|^2}{s^\prime} + |z^\prime_{0}| \varpi \right ) \exp ( - |z^\prime_{0}| \varpi ) \right ]
\nonumber\\
&+ \frac{1}{2 {s^\prime}^2 (z^\prime_{2})^3} \left[
\left ( 1 + \frac{4|b^\prime| (z^\prime_{2})^2}{ s^\prime} \right ) \sin (z^\prime_2 \varpi) - z^\prime_{2}{\varpi} \cos (z^\prime_2 \varpi)
\right ]
\label{eq:PSDFcaseII}
\end{align}
for $\varpi > 0$. 
In this order of approximation, the envelope of $\tilde{g}^{\prime}_U(\varpi)$ acquires its typical inverted-S shape when $0<\kappa^\prime \ll 1$, as observed in \cite[Fig. 1]{arnalocavg}, which was based instead on a higher-order exponential curve fit. 

For $0<\kappa^\prime \ll 1/4$, eq. (\ref{eq:PSDFcaseII}) can be expressed as
\begin{align}
&\tilde{g}^{\prime}_U(\varpi) 
\simeq \frac{1}{\sqrt{2 \lambda^\prime_2}} \nonumber\\
&\times \left [ 1 + \frac{3\kappa^\prime}{2} + \left ( 1 - 3 \kappa^\prime \right ) \sqrt{\frac{2}{\lambda^\prime_2}} \varpi \right ] 
\exp \left ( - \sqrt{(1-{\kappa^\prime})\frac{2}{\lambda^\prime_2}} \, \varpi \right ) 
\nonumber\\ 
&+ 
5 \sqrt{\frac{{\kappa^\prime}^3}{2 \lambda^\prime_2}} 
\sin \left ( \sqrt{\frac{2}{{\kappa^\prime} \lambda^\prime_2 }} \,\varpi \right ) 
- \frac{\kappa^\prime}{\lambda^\prime_2} \varpi
\cos \left ( \sqrt{\frac{2}{{\kappa^\prime} \lambda^\prime_2 }} \,\varpi \right ) 
\label{eq:PSDFcaseII_smallkappa}
\end{align}
reducing to (\ref{eq:psdf_nonexp_U}) in the limit $\kappa^\prime \rightarrow 0$. 

\begin{figure}[htb] \begin{center}
\begin{tabular}{c}
\vspace{-0.5cm}\\
\hspace{-0.7cm}
\includegraphics[scale=0.66]{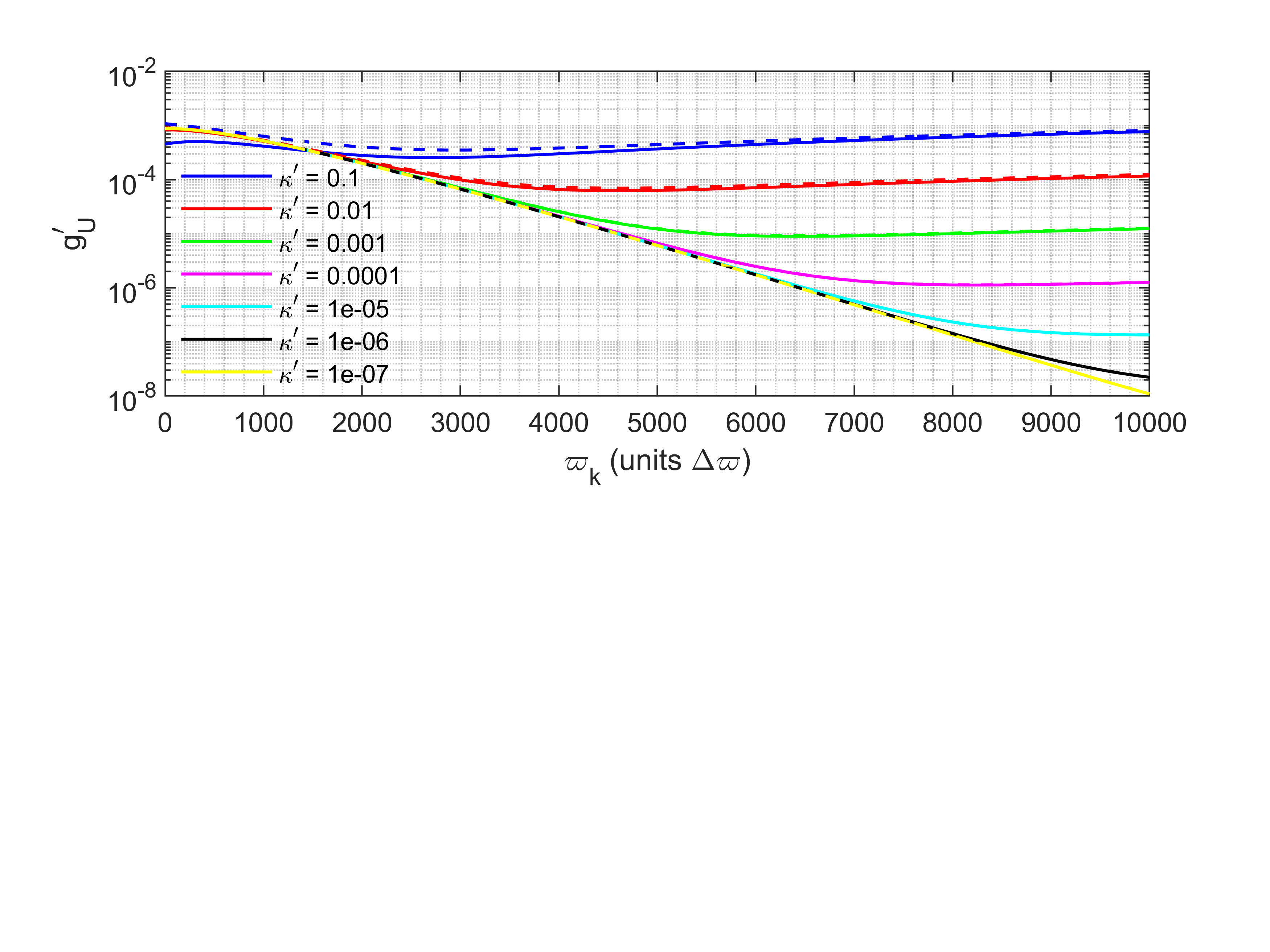}\\
\vspace{-4.5cm}\\
\\
\end{tabular}
\end{center}
\caption{
\label{fig:psdf_theo}
\small
PSDF $\tilde{g}^\prime_U(\varpi)$ for $[0/2]$-order Pad\'{e} approximation based ACF $\tilde{\rho}^\prime_U(\tau)$ at selected values of $\kappa^\prime$, for 
$(\lambda^\prime_2)^{1/2} = 1000 $ rad/s and $\Delta \varpi = 0.01$ rad/s. Solid: envelope of (\ref{eq:PSDFcaseII}) for general $\kappa^\prime$; dashed: envelope of small-$\kappa^\prime$ approximation (\ref{eq:PSDFcaseII_smallkappa}), 
reducing to (\ref{eq:psdf_nonexp_U}) for $\kappa^\prime \rightarrow 0$.
}
\end{figure}

The asymptotic envelopes of (\ref{eq:PSDFcaseII}) and (\ref{eq:PSDFcaseII_smallkappa}), obtained by bounding $\sin(\cdot)$ and $-\cos(\cdot)$ by $1$, are shown in Fig. \ref{fig:psdf_theo} for selected values of $\kappa^\prime$ with $\sqrt{\lambda^\prime_2} = 1000$ rad/s. If $\kappa^\prime\rightarrow 0$ then 
$\tilde{g}^\prime_U(\varpi)$ acquires an inverted-S shape, a mean decay rate of $\sqrt{2/\lambda^\prime_2}$, a higher value of $\varpi^\prime_{c,U}$, and a flattening of the stir HF part where only the dominant first term in (\ref{eq:PSDFcaseII}) or (\ref{eq:PSDFcaseII_smallkappa}) remains. It then also approaches (\ref{eq:psdf_nonexp_U}). This first-order model already exhibits flattening at low stir frequencies, but with an unrestricted exponential decay at high stir frequencies, in contrast to the second-order model. 
Both LF and HF flattening on average were observed experimentally in \cite[Fig. 1]{arnalocavg}.

In summary, the doubling of the multiplicity of the poles in the $[0/2]$-order Pad\'{e} approximated ACF $\tilde{\rho}^\prime_U(\tau)$ in (\ref{eq:rhoU_expanded_Pade_2ndorder}), compared to $\tilde{\rho}^\prime_E(\tau)$ in (\ref{eq:rhoE_expanded_Pade_2ndorder}), has the effect of flattening the SDF, both near stir DC (observable in both the first- and second-order approximations of $g_U(\varpi)$) and at high stir frequencies (in the second-order approximation only). 
In the stir domain, HF flattening represents whitening and yields a more narrower ${\rho}^\prime_U(\tau)$ around $\tau=0$ compared to ${\rho}^\prime_E(\tau)$.

\section{ACFs and SDFs for Additive White Noise\label{sec:noise}}
The foregoing models were based on an indiscriminate cause of the HF departure from $g^\prime_E(\varpi)=\beta^\prime\exp(-\beta^\prime \varpi)$. 
We shall further drop the primes for simplicity.
When stir noise or electrical noise are separable from the stirred field, an extended model and representation can be obtained by superposition of their non-normalized ASDFs, resulting in a composite ASDF.
In practice, stirred fields and their measurement are typically affected by ambient EMI and various sources of noise \cite{midd}. 

Let the observed field $E(t)$ be the linear superposition of a noiseless stirred 
excitation field $E_0(t)$ and a noise field $N(t)$ of unspecified origin at an observation time $t$, i.e.,
\begin{align}
E(t) = E_0 (t) + N(t)
.
\label{eq:noise_model}
\end{align} 
Both $E_0(t)$ and $N(t)$ are assumed to be WSS, independent, circular random fields of unlimited bandwidth  with respective variances $\sigma^2_{E_0}$ and $\sigma^2_N$. 
Their ACFs $\rho_{E_0}(\tau)$ and $\rho_N(\tau)$ may exhibit different functional forms and correlation lengths, $\tau_{\rmc, E_0}$ and $\tau_{\rmc, N}$. Substituting (\ref{eq:noise_model}) into (\ref{eq:rho_gen}) yields $\rho_E(\tau)$ as a variance-weighted sum of $\rho_{E_0}(\tau)$ and $\rho_N(\tau)$, i.e.,
\begin{align}
\rho_E(\tau) 
&= \frac{\sigma^2_{E_0} \rho_{E_0}(\tau) + \sigma^2_N \rho_N(\tau)}{\sigma^2_{E_0} + \sigma^2_{N} }\label{eq:ACF_noisymeas_basic}\\
&= \rho_{E_0} (\tau) 
\frac{1 + (\sigma^2_N/\sigma^2_{E_0}) \rho_N(\tau)/\rho_{E_0} (\tau) }{1 + (\sigma^2_N/\sigma^2_{E_0})}
\label{eq:ACF_noisymeas}
\end{align}
which is a function of the stir-to-noise ratio (SNR) of their RMS levels, $\sigma^2_{E_0}/\sigma^2_N$, and the correlation ratio $\rho_{E_0}(\tau)/\rho_N(\tau)$ at equal stir and noise delay, $\tau$. 
For relatively high SNR levels ($\sigma^2_N/\sigma^2_{E_0} \ll 1$), eq. 
(\ref{eq:ACF_noisymeas}) is approximated as
\begin{align}
\rho_E(\tau) \simeq \rho_{E_0} (\tau) -\frac{\sigma^2_N}{\sigma^2_{E_0}} \left [ \rho_{E_0} (\tau) - \rho_{N} (\tau) \right ]
.
\label{eq:ACF_E_noisymeas_smallNSR}
\end{align} 
Typically, $\tau_{\rmc, N} \ll \tau_{\rmc, E_0}$ and $|\rho_{N} (\tau) | \leq |\rho_{E_0} (\tau)|$, which results in (\ref{eq:ACF_E_noisymeas_smallNSR}) exhibiting a reduced ACF value of $E$ compared to $E_0$.
The associated ASDF of $E$ then follows from  (\ref{eq:ACF_noisymeas_basic}) as
\begin{align}
g_E(\varpi) 
&\simeq \left ( 1 - \frac{\sigma^2_N}{\sigma^2_{E_0}} \right ) g_{E_0}(\varpi) + \frac{\sigma^2_N}{\sigma^2_{E_0}} g_{N}(\varpi),\hspace{3mm}\frac{\sigma^2_N}{\sigma^2_{E_0}} \ll 1
.
\label{eq:SDF_noisymeas_smallNSR}
\end{align} 
In particular, (\ref{eq:SDF_noisymeas_smallNSR}) shows that weak additive white noise, i.e., 
$\rho_N(\tau) = \delta(\tau)$ and $g_N(\varpi) = 1$, gives rise to a constant noise floor $\sigma^2_N/\sigma^2_{E_0}$ for $g_E(\varpi)$ when $\varpi \rightarrow +\infty$. 
It also reduces $g_{E_0}(\varpi)$  
by a factor $1 - {\sigma^2_N}/{\sigma^2_{E_0}}$. Evaluation of the HF stir noise floor level can be used to adjust the slope of the SDF to infer the true $g_{E_0}(\varpi)$ if noise cannot be isolated from $E_0$.

For the corresponding energy $U$, starting from noiseless circular $E_0$, i.e., $\rho_{|E_0|^2} \simeq \rho^2_{E_0}$, it follows from (\ref{eq:ACF_noisymeas}) that
\begin{align}
\rho_U (\tau) \simeq 
\left ( 1 - 2 \frac{\sigma^2_N}{\sigma^2_{E_0}} \right ) \rho^2_{E_0} (\tau) + 2 \frac{\sigma^2_N}{\sigma^2_{E_0}} \rho_{E_0} (\tau) \rho_{N} (\tau) 
\label{eq:ACF_U_noisymeas_smallNSR}
\end{align}
whence the noise reduces the stir content in the PSDF, viz., 
\begin{align}
g_U (\varpi) \simeq 
\left ( 1 - 2 \frac{\sigma^2_N}{\sigma^2_{E_0}} \right ) g_{U_0} (\varpi) 
+ 2 \frac{\sigma^2_N}{\sigma^2_{E_0}} 
g_{E_0}(\varpi)* g_N(\varpi)
\label{eq:SDF_U_noisymeas_smallNSR}
\end{align}
in which $*$ represents linear convolution.

\section{Conclusion}
In this part I, auto- and cross-spectral density function models for fields and power were obtained based on Pad\'{e} approximants for their corresponding correlation function models. 
These functions depend solely on the normalized spectral moments $\lambda^{\prime(\prime)}_m$ as parameters, which can be independently evaluated from stir sweep data. 
The approximants reveal the presence and extent of stir noise, whose effect can be expressed via a hierarchy of stir spectral distortion coefficients $\kappa^{\prime(\prime)}_m$. 
Depending on the particular value of $\kappa^{\prime(\prime)} \equiv \kappa^{\prime(\prime)}_2$, relatively simple explicit expressions of the SDFs were obtained.

A further model was presented in which the overall ASDF is explicitly represented as a sum of individual ASDFs for a stir process and a noise process. 
Both methods need not be mutually exclusive: a Pad\'{e} approximation can be applied to a composite spectral process before adding additional sources of noise. Conversely, a Pad\'{e} approximant could be further decomposed into a set of elementary ASDF components.

The analysis started from a serial expansion and approximation of CFs in terms of $\lambda^{\prime(\prime)}_m$, with the SDF following by Fourier transformation. 
Conversely, one may start by expanding a continuous ensemble or sample ASDF expressed in terms of the autocorrelation moments 
\begin{align}
\theta^{\prime}_n =
\int^{+\infty}_{-\infty} |\tau|^{n} \rho^{\prime}_E(\tau) \rmd \tau,\hspace{3mm}
n=0,1,2,\ldots
\end{align}
The expanded ASDF can then be assigned its own (stir spectral) Pad\'{e} approximation $\tilde{g}^{\prime}_E(\varpi)$, followed by an inverse Fourier cosine transform to $\tilde{\rho}^{\prime}_E(\tau)$. 
A similar approach can be applied to extracting a sample CCF from a series expansion of a CSDF.


\begin{thebibliography}{99}
\bibitem{arnalocavg} L. R. Arnaut, ``Effect of local stir and spatial averaging on the measurement and testing in mode-tuned and mode-stirred reverberation chambers,'' \it IEEE Trans. Electromagn. Compat., \rm vol. 43, no. 3, pp. 305--325, Aug. 2001.
\bibitem{well2001} N. Wellander, O. Lund\'{e}n, and M. B\"{a}ckstr\"{o}m, ``The maximum value distribution in a reverberation chamber,'' \it Proc. IEEE Symp. Electromagn. Compat., \rm Montr\'{e}al, Canada, Aug. 2001, pp. 751--756.
\bibitem{lemo2008} C. Lemoine, P. Besnier, and M. Drissi, ``Estimating the effective sample size to select independent measurements in a reverberation chamber,'' \it IEEE Trans. Electromagn. Compat., \rm vol. 50, no. 2, pp. 227--236, May 2008.
\bibitem{pirk2012} R. J. Pirkl, K. A. Remley, and C. S. L. Patan\'{e}, ``Reverberation chamber measurement correlation,'' \it IEEE Trans. Electromagn. Compat., \rm vol. 53, no. 3, pp. 533--545, Jun. 2012.  
\bibitem{mons2015} F. Monsef, R. Serra, and A. Cozza, ``Goodness-of-fit tests in reverberation chambers: is sample independence necessary?'', \it IEEE Trans. Electromagn. Compat., \rm vol. 57, no. 6, pp. 1748--1751, Dec. 2015.
\bibitem{arnaexcessletter} L. R. Arnaut, ``Excess and deficiency of extreme multidimensional random fields,'' \it IEEE Trans. Electromagn. Compat., \rm vol. 64, no. 1, pp. 255--258, Feb. 2022.
\bibitem{arnathresh} L. R. Arnaut, ``Threshold level crossings, excursions, and extrema in immunity and fading testing using multistirred reverberation chambers,'' \it IEEE Trans. Electromagn. Compat., \rm vol. 62, no. 5, pp. 1638--1650, Oct. 2020.
\bibitem{arnaexcess} L. R. Arnaut, ``Excess power, energy, and intensity of stochastic fields in quasi-static and dynamic environments,'' \it IEEE Trans. Electromagn. Compat., \rm vol. 63, no. 3, pp. 792--802, Jun. 2021.
\bibitem{caro2022} C. F. M. Carobbi and R. Serra, ``Exponential correlation model for field intensity in reverberation chambers,'' \it Proc. 16th Eur. Conf. Ant. Propag. (EuCAP), \rm Madrid, Spain, 27 Mar. - 1 Apr. 2022.
\bibitem{hill1995} D. A. Hill, ``Spatial correlation function for fields in a reverberation chamber,'' \it IEEE Trans. Electromagn. Compat., \rm vol. 37, no. 1, p. 138, Feb. 1995.
\bibitem{hillladbury} D. A. Hill and J. M. Ladbury, ``Spatial-correlation functions of fields and energy density in a reverberation chamber,''  \it IEEE Trans. Electromagn. Compat., \rm vol. 44, no. 1, pp. 95--101, Feb. 2002.
\bibitem{arnaPRE} L. R. Arnaut, ``Spatial correlation functions of inhomogeneous random electromagnetic fields,''  \it Phys. Rev. E, \rm vol. 73, no. 3,  036604, Apr. 2006.
\bibitem{arnacopula} L. R. Arnaut, ``Copulas, outliers, and rogue states of nonelliptic fields and energy in electromagnetic reverberation,'' \it IEEE Trans. Electromagn. Compat., \rm vol. 58, no. 2, pp. 371--385, Apr. 2016.
\bibitem{xu2018} Q. Xu, L. Xing, Y. Zhao, Z. Tian, and Y. Huang, ``Wiener--Khinchin theorem in a reverberation chamber,'' \it IEEE Trans. Electromagn. Compat., \rm vol. 61, no. 5, pp. 1399--1407, Oct. 2019.
\bibitem{arnapulsejitt} L. R. Arnaut, ``Pulse jitter, delay spread, and Doppler shift in mode-stirred reverberation,'' \it IEEE Trans. Electromagn. Compat., \rm vol. 58, no. 6, pp. 1717--1727, Dec. 2016.
\bibitem{wong1985} E. Wong and B. Hajek, \it Stochastic Processes in Engineering Systems. \rm Springer: New York, NY, USA, 1985.
\bibitem{arnaellipt} L. R. Arnaut, ``Elliptic stochastic fields in reverberation chambers,'' \it IEEE Trans. Electromagn. Compat., \rm vol. 58, no. 1, pp. 11--21, Feb. 2016.
\bibitem{vanm1983} E. Vanmarcke, \it Random Fields: Analysis and Synthesis. \rm 2nd ed. World Scientific: Singapore, 2010.
\bibitem{kac1959} M. Kac and D. Slepian, ``Large excursions of Gaussian processes,'' \it Ann. Math. Statist., \rm vol. 30, pp. 1215--1228, 1959.
\bibitem{brac1986} R. N. Bracewell, \it The Fourier Transform and Its Applications. \rm 2nd ed., revised. McGraw--Hill: New York, NY, 1986.
\bibitem{bake1996} G. A. Baker (Jr.) and P. R. Graves-Morris, \it Pad\'{e} Approximants. \rm 2nd ed. Cambridge Univ. Press: Cambridge, U.K., 1996.
\bibitem{cram1946} H. Cram\'{e}r, \it Mathematical Methods in Statistics. \rm Princeton Univ. Press: Princeton, NJ, 1946.
\bibitem{bril1981} D. R. Brillinger, \it Time Series: Data Analysis and Theory. \rm Holt, Rinehart and Winston: New York, NY, 1975.
\bibitem{bell1965} P. A. Bello, ``On the rms bandwidth of nonlinearly envelope detected narrow-band Gaussian noise,'' \it IEEE Trans. Inf. Theo., \rm vol. 11, no. 2, pp. 236--239, Apr. 1965.
\bibitem{midd} D. Middleton, \it An Introduction to Statistical Communication Theory. \rm McGraw--Hill: New York, NY, 1960.
\bibitem{grad} I. S. Gradshteyn and I. M. Ryzhik, \it Table of Integrals, Series, and Products. \rm 7th ed. Academic: London, U.K., 2007. 
\end{thebibliography}
\end{document}